\documentclass[prc,twocolumn,showpacs,showkeys,preprintnumbers,floatfix,nofootinbib,superscriptaddress]{revtex4-1}
\usepackage{graphicx,amsmath,amssymb,bm,color,inputenc,fontenc,mathrsfs}
\usepackage{epstopdf}
\usepackage{float}
\usepackage[caption = false]{subfig}
\usepackage{comment}
\usepackage[normalem]{ulem}
\usepackage{soul}
\usepackage{bbold}
\usepackage{hyperref} % hypertext links 
\usepackage{bm} % ----------------------------------------------------------------

\newcommand{\be}{\begin{equation}}
\newcommand{\ee}{\end{equation}}
\newcommand{\bea}{\begin{eqnarray}}
\newcommand{\eea}{\end{eqnarray}}

\def\bea{\begin{eqnarray}}
\def\eea{\end{eqnarray}}
\newcommand{\beq}{\begin{equation}}
\newcommand{\eeq}{\end{equation}}
\newcommand{\beast}{\begin{equation*}}
\newcommand{\eeast}{\end{equation*}}

\newcommand\ket[1]{| #1 \rangle}

\usepackage[dvipsnames,usenames]{xcolor}

\begin{document}

\title[title]{Exact representations of many-body interactions with RBM neural networks}% and applications in Quantum Annealers}
\author{Ermal Rrapaj}
 \email{ermalrrapaj@gmail.com}
\affiliation{Department of Physics, University of California, Berkeley, CA 94720, USA}
 \affiliation{School of Physics and Astronomy, University of Minnesota, Minneapolis, MN 55455, USA}
 
 \author{Alessandro Roggero}
 \email{roggero@uw.edu}
\affiliation{Institute for Nuclear Theory, University of Washington, 
Seattle, WA 98195, USA}

\preprint{INT-PUB-20-016}
\date{\today}

\begin{abstract}
Restricted Boltzmann machines (RBMs) are simple statistical models defined on a bipartite graph which have been successfully used in studying more complicated many-body systems, both classical and quantum. In this work, we exploit the representation power of RBMs to provide an exact decomposition of many-body contact interactions into one-body operators coupled to discrete auxiliary fields. This construction generalizes the well known Hirsch's transform used for the Hubbard model to more complicated theories such as pionless effective field theory in nuclear physics, which we analyze in detail. We also discuss possible applications of our mapping for quantum annealing applications and conclude with some implications for RBM parameter optimization through machine learning.
\end{abstract}

\maketitle

\section{Introduction}
In recent decades, statistical methods based on artificial neural networks (ANN) have been proven extremely valuable in studying the physical world. For a long time, experimental high energy physics has been at the forefront of these applications~\cite{Jobes1972,Olmos1991,Denby1999} and today ANN-based methods are of fundamental importance in analyzing particle accelerator data~\cite{Ciodaro2012,Edelen2016,Radovic2018,Scheinker2018}. Thanks to the growing availability of large scale computational resources, these types of approaches have started to play an important role also in many-body theory more generally with applications as diverse as detecting phase transitions in simulations~\cite{Lei2016,Carrasquilla2017}, preparing accurate variational state for lattice systems~\cite{Carleo2016}, accelerating sampling in Monte Carlo based simulations~\cite{Huang2017,Liu2017,Liu2018}, constructing efficient energy density functionals~\cite{Nelson2019} and performing fast approximate quantum state tomography~\cite{Torlai2018,Carrasquilla2019}.

A particularly interesting class is the generative models commonly used in unsupervised learning whose aim is to automatically discover underlying patterns in the data that are being analyzed (see eg.~\cite{Bengio2013,MEHTA2019} for an introduction). The main advantage of these class of methods is the possibility of finding a compact and possibly accurate description of the supplied data without using any pre-determined labeling. On one side, this allows the automatic discovery of appropriate labels and on the other simplifies dramatically the process of acquiring useful data points to train the model on.
Once model has been trained it can be used, for instance, to generate new data points from the reconstructed density distribution~\cite{MEHTA2019}.

Our focus in this paper is a particularly simple ANN model composed by only two layers of neurons with a very sparse connectivity: the restricted Boltzmann machine (RBM)~\cite{Smolensky1986, Freund1992}. We will describe this architecture in some detail in Sec.~\ref{sec:conjecture}, but for now we want to anticipate that an RBM can be used as a universal approximator for arbitrary probability distributions (see eg.~\cite{LeRoux2008,Montufar2011}) . This fact, together with it's simplicity, is one of the main reasons for it's wide-spread use in many-body physics applications~\cite{Carleo2016,Huang2017,Carrasquilla2017}. We note in passing that this construction is also the basic building block of deep learning models including the deep Boltzmann machine (DBM) and deep belief network (DBN)~\cite{Hinton2006,Salakhutdinov2009,Montufar2018} where the number of layers is increased to provide more flexibility to the representation power of the model. For instance, three layer DBMs have been successfully used in exact representations of ground state wavefunctions~\cite{Carleo2018}.

In this work we will use an RBM model to find simpler but exact representations for the many-body partition function
\begin{equation}
 \exp\left(-\beta \hat{H}\right) \propto \text{Tr}_{\bf{s}}\left[\exp\left(-F_{\text{rbm}}\left(\{\hat{\rho}_i\},\bf{s}\right)\right)\right]\;,
\label{eq:zbeta}
\end{equation}
in terms of the free energy $F_{\text{rbm}}$ of an RBM where the visible layer is composed by quantum operators $\{\hat{\rho}_i\}$ and the hidden layer is made by a vector of classical auxiliary fields $\bf{s}$ which we marginalize over. Exact representations of this form are of fundamental importance for Quantum Monte Carlo (QMC) calculations of many-body systems~\cite{Foulkes2001,Carlson2015} which approximate ground-state expectation values as
\begin{equation}
\langle 0 \lvert \hat{O}\rvert0\rangle = \lim_{\beta\to\infty} \frac{1}{Z(\beta)} Tr\left[\exp\left(-\beta \hat{H}\right) \hat{O}\right]\;,
\end{equation}
where we introduced the partition function $Z(\beta)=\text{Tr}\left[e^{-\beta\hat{H}}\right]$.
More specifically, given a many-body Hamiltonian $\hat{H}$ written as a linear combination of Hermitian operators $\hat{H} = \sum_k \hat{H}_k$ and some initial state $\ket{\Psi}$, the basic computational step needed for QMC simulations is the map
\begin{equation}
\ket{\Psi}\longrightarrow \ket{\Psi_k} = \exp\left(-\beta \hat{H}_k\right) \ket{\Psi}\;,
\end{equation}
for each one of the $\hat{H}_k$ terms composing the Hamiltonian operator $\hat{H}$. Efficient schemes to simplify the evolution operators $\exp\left(-\beta \hat{H}_k\right)$ using a map like Eq.~\eqref{eq:zbeta} are key ingredients of QMC methods. Notable examples include auxiliary field methods such as the Hubbard-Stratonovich~\cite{Stratonovich1957,Hubbard1959} and the Hirsch's transform~\cite{Hirsch1983} or the more recent approach proposed by K{\"o}rber, Berkowitz and Luu in~\cite{Korber2017}.

In this work we focus in particular to two classes of Hamiltonians of fundamental importance in nuclear physics:
\begin{itemize}
    \item realistic local interactions~\cite{wiringa1995,Gezerlis2014} represented by interactions of the form
    \begin{equation}
    \label{eq:vlocal}
    \hat{H}_V = \sum_i V_i(\vec{R}) w_i(\vec{\sigma},\vec{\tau})\;,
    \end{equation}
    with $V_i(\vec{R})$ scalar coefficients dependent on the nucleon coordinates $\vec{R}$ and $w_i(\vec{\sigma},\vec{\tau})$ a functional of spin operators $\vec{\sigma}$ and iso-spin operators $\tau$. Note that it is always possible to choose the operators $w_i$ in Eq.~\eqref{eq:vlocal} to be involutive: $w_i(\vec{\sigma},\vec{\tau})^2 = \mathbb{1}$.
    \item nuclear potentials derived in low energy effective theories expressed in terms of two and three-body contact interaction~\cite{Lee2009} of the form
    \begin{equation}
    \label{eq:veft}
    \begin{split}
    \hat{H}_V &= \sum_{i,j} v_{ij} \hat{\rho}_i\hat{\rho}_j w_{ij} (\vec{\sigma},\vec{\tau})\\
    &+\sum_{i,j,k} v_{ijk} \hat{\rho}_i\hat{\rho}_j\hat{\rho}_k w_{ijk} (\vec{\sigma},\vec{\tau})\;.
    \end{split}
    \end{equation}
    In this expression $v_{ij}$ and $v_{ijk}$ are scalar coefficient, $\hat{\rho}_i$ is a fermionic density operator (see Eq.~\eqref{eq:fdens} below for a more formal definition) and $w_{ij}(\vec{\sigma},\vec{\tau})$ and $w_{ijk}(\vec{\sigma},\vec{\tau})$ are idempotent spin and iso-spin operators as in Eq.~\eqref{eq:vlocal}. We note that these interaction arise naturally also in the low energy description of condensed matter systems (see eg.~\cite{Nishida2010,Carlson2017}).
\end{itemize}
Thanks to this representation it is sufficient to find an exact mapping in Eq.~\eqref{eq:zbeta} for idempotent operators only, and this will be the focus of our present work. Note that it is always possible to express any operator on a finite Hilbert space as a linear combination of involutive operators by using (tensor product of) Pauli operators as an operator basis.

In the rest of the paper, we provide an introduction to the RBM and present it's application to represent many-body forces in nuclear physics. In section~\ref{sec:conjecture} we proceed to relate the free energy $F_{\text{rbm}}$ of this architecture to the physical partition functions produced by many-body interactions, starting with the familiar case of a two-body potential term. We then present a novel generalization to the case of three-body forces in Sec~\ref{sec:3bd}. Further details on the construction for general terms is provided in Appendix~\ref{app:General_EQ} for completeness.
One of the advantages of our approach is that it can be easily generalized from binary auxiliary fields to generic categorical classical variables which take values on a larger set $\{0,1,\dots,\mathcal{K}-1\}$; in Sec~\ref{sec:categorical} we discuss the advantages that this added flexibility can provide.

We then proceed in Sec.~\ref{sec:qa} to show how this RBM mapping could be used to improve the representation power of quantum annealers based on the transverse Ising model Hamiltonian. In Sec.~\ref{sec:train} we use the exact mapping obtained before to analyze numerically the performance of various optimization protocols for the RBM parameters in reaching the known optimum for the simple case of a 2D Ising model. Note that this optimization step is not directly needed to use our results but instead addresses the question of feasibility of machine learning through RBMs and is of independent interest.

We conclude in Sec.~\ref{sec:conclusion} with a summary and possible implications of our results.

\section{Restricted Boltzmann machines as Hamiltonians with auxiliary fields}
\label{sec:conjecture}
The introduction of auxiliary fields as a mean of simplifying the interaction term
of the Hamiltonian is common practice in many areas of theoretical physics, and is 
particularly popular in designing Quantum  Monte Carlo algorithms in
both condensed matter and nuclear physics~\cite{Blankenbecler1981,Hirsch1985,Alhassid1994,SUGIYAMA1986,Johnson1992,Zhang1995,Foulkes2001,Carlson2015}.
In these applications, a system of interacting particles is mapped into a free theory
coupled to a background fluctuating auxiliary field and the final simulations
is usually performed after integrating out the physical fields.

One famous instance of this class of mappings is the Hubbard-Stratonovich transformation~\cite{Stratonovich1957,Hubbard1959} which exploits the Gaussian-integral relation
\begin{equation}
 \begin{split}
  \exp \left(\frac{\tau}{2}\hat{O}^2\right)=\frac{1}{\sqrt{2 \pi}}\int_{-\infty}^\infty dh e^{-h^2/2-\sqrt{\tau} h \hat{O}},
 \end{split}
 \label{eq:evolve2bd}
\end{equation}
with $h$ a real auxiliary field, to provide a simpler representation for the evolution operator on the left hand side. This transformation is used extensively to express the evolution under two-body interactions as a superposition of evolutions under one-body interactions parameterized by the value of the auxiliary field and is commonly used in Auxiliary Field Monte Carlo techniques~\cite{SUGIYAMA1986,Johnson1992,Zhang1995,Carlson2015}. This transformation can also be used for higher-order operators when they can be expressed as squares, a common example is the three-neutron force (see eg.~\cite{Gandolfi2009}). When the interaction cannot be written as a perfect square one can attempt a recursive application of the transformation in Eq.~\eqref{eq:evolve2bd} but this is usually accompanied by a drastically reduced efficiency (for an example of this applied to the isospin-dependent spin-orbit force see~\cite{Zhang2014}).

\begin{figure}[ht]
 \includegraphics[scale=0.45]{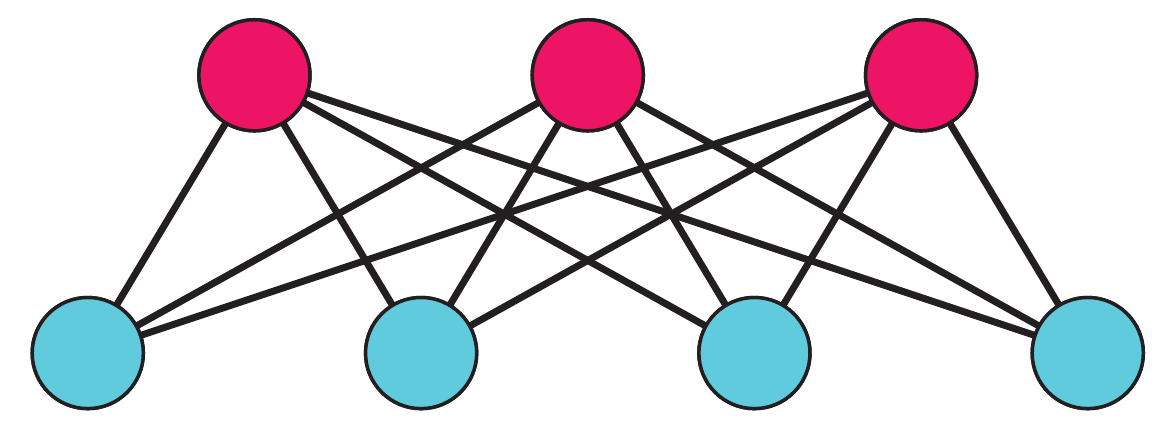}
 \caption{Restricted Boltzmann machine, visible layer below (in blue) and hidden layer above (in red).}
 \label{fig:rbm}
\end{figure}

Another popular family of transformations that achieve a similar simplification can be obtained by considering auxiliary fields which can only take a discrete number of values. The prototypical example of this is the famous Hirsch transformation for the Hubbard model~\cite{Hirsch1983} which we will discuss in some detail in the next subsection (see also~\cite{Assaad1997,Motome1997} for similar constructions). The main purpose of our work is to show how these discrete mappings can be described and generalized using the language of Restricted Boltzmann Machines (RBM)~\cite{Smolensky1986, Freund1992}. 
As we briefly described in the introduction and show pictorially in Fig.~\ref{fig:rbm}, the RBM is a statistical model organized in two interconnected layers: a {\it visible layer} composed by a vector of $N_v$ visible units $\bf{v}$ (depicted as blue dots in Fig.~\ref{fig:rbm}) which represent the dynamical variables whose statistics we want to model, and a {\it hidden layer} composed by a vector of $N_h$ hidden units $\bf{h}$ (depicted as red dots in Fig.~\ref{fig:rbm}) which represent auxiliary variables used to create correlations among the visible units. The RBM is an energy based statistical model, meaning that the probability distribution associated with the network can be conveniently expressed as a partition function 
\begin{equation}
\label{eq:rbmprob}
 P_{\text{rbm}}\left(\bf{v},\bf{h}\right) = \mathcal{N}\exp\left(-F_{\text{rbm}}\left(\bf{v},\bf{h}\right)\right)
\end{equation}
with $\mathcal{N}$ a normalization constant and free energy
\begin{equation}
 F_{\text{rbm}}\left(\bf{v},\bf{h}\right) = {\bf B} \cdot {\bf v}+ {\bf C}\cdot{\bf h}+\sum_{i=1}^{N_v}\sum_{j=1}^{N_h}W_{ij}v_ih_j\;.
\end{equation}
This free energy is parameterized by the two {\it bias} vectors $\bf{B}$ and $\bf{C}$, which independently shift the probability density of the variables in the two layers, and the {\it weight} matrix $W$ which couples the two layers. In Fig.~\ref{fig:rbm} the latter is represented by the black connections. The reason for this choice as a network structure is linked to the conditional independence of the variables on the two layers which allows for an efficient sampling of the conditional probabilities
\begin{equation}
p_v\left(\bf{v}\right)= P_{\text{rbm}}\left(\bf{v}|\bf{h}\right)\;\;\text{and}\;\; p_h\left(\bf{h}\right)=P_{\text{rbm}}\left(\bf{h}|\bf{v}\right)\;,
\end{equation}
which can then be used to draw efficiently samples from the wanted marginal probability distribution
\begin{equation}
P^v_{\text{rbm}}\left(\bf{v}\right) = \text{Tr}_{\bf{h}}\left[ P_{\text{rbm}}\left(\bf{v},\bf{h}\right)\right]\;,
\end{equation}
using block Gibbs sampling. In the expression for the marginal probability $P^v_{\text{rbm}}$ above, we used Tr$_{\bf{h}}$ to denote a summation over all the possible values of the hidden variables. As we mentioned briefly in the introduction, despite its simplicity the RBM is a universal approximator~\cite{LeRoux2008,Montufar2011}, in the sense that by adding a sufficiently large number of hidden variable one can accurately approximate any marginal probability distribution $P^v_{\text{rbm}}$ of the visible variables.

By promoting the units on which the RBM is defined to operators acting on some Hilbert space one can also define a Quantum RBM (see eg.~\cite{Amin2018}) allowing statistical inference on quantum states. We will defer a more in-depth discussion about this model, and it's possible implementation using quantum annealers to Sec.~\ref{sec:qa}.

The main contribution of our work is the use of the RBM network structure to define a hybrid quantum-classical architecture, where the visible unit is composed by quantum operators and the hidden units are classical discrete variables, to generalize the auxiliary-field decomposition of interactions to arbitrary many-body forces.
To the best of our knowledge the closest construction to our architecture is the recently proposed map between generalized Ising models and Deep Boltzmann Machines presented in~\cite{Yoshioka2019}. Our method overcomes some of the difficulties encountered in that proposal by reducing the depth of the required network (and thus simplifying sampling) and removing any restriction on the numerical value of the coupling constants.

To proceed further and present the model in more details, we will now focus on the description of physical systems containing $N_f$ species~\footnote{for instance $N_f=2$ for neutrons due to the 2 possible spin projections} of interacting fermions and discretized over $N$ modes. 
General contact interactions can be expressed as powers of the 
 density operator $\hat{\rho}_a(k)$ for mode $k$ and species $a$ as
\begin{equation}
\label{eq:fdens}
\hat{\rho}_a(k) = c^\dagger_a(k)c_a(k)\quad\{c^\dagger_a(k),c_b(q)\} = \delta_{k,q}\delta_{a,b}\;,
\end{equation}
with $c^\dagger_a(k)$ and $c_a(k)$ fermionic creation and annihilation operators and $\{\cdot,\cdot\}$ the anti-commutator. For instance 2 and 3 body contact interactions can be written as
\begin{equation}
\hat{V}_2 = \hat{\rho}_a(k) \hat{\rho}_b(k)\quad\hat{V}_3 = \hat{\rho}_a(k) \hat{\rho}_b(k)\hat{\rho}_c(k)\;,
\end{equation}
% As we will show in more detail in Sec.~\ref{sec:general_int} our construction can be generalized to more general interactions..
Note that, due to the Pauli exclusion principle, identical fermions cannot occupy the same quantum state within the system. Thus, $N_f$ species in a system allows for many-body forces of up to $N_f$ for a given mode. In order to simplify the notation we will use the multi index $\mu=(a,k)$, taking values from $1$ to $M=N_fN$, and use the short-hand $\hat{\rho}_\mu$ to indicate the density operator $\hat{\rho}_a(k)$.

Using these density operators as our visible units, we can now write the free energy of our hybrid classical-quantum model as
 \begin{equation}
 \label{eq:RBM_E1}
 \begin{split}
F_{\text{rbm}}&\left(\hat{\bm{\rho}},\bf{h}\right) = {\bf B}\cdot\hat{\bm{\rho}} + {\bf C}\cdot{\bf h}+\sum_{\mu=1}^{M}\sum_{j=1}^{N_h}W_{ij}\hat{\rho}_\mu h_j\;.
% &=\sum_{\mu=1}^{M} B_\mu\hat{{\rho}}_\mu + \sum_{j=1}^{N_h} C_j h_j+\sum_{\mu=1}^{M}\sum_{j=1}^{N_h}W_{ij}\hat{\rho}_\mu h_j\;.
 \end{split}
\end{equation}%\ale{**Check if the second line belongs here after rereading the whole thing**}
In order to simplify the exposition we will consider for the moment the special case where the hidden units are binary variables $h_j=\{0,1\}$, and generalize the construction to more general categorical variables in Sec.~\ref{sec:categorical}. 

Before describing in detail the special case of 2 and 3 body forces, we want now to show how, by carefully choosing the parameters that define the free energy Eq.~\eqref{eq:RBM_E1}, we can obtain all the possible contact interactions up to the maximum order $M$. By tracing out the hidden layer from the total probability distribution $P_\text{rbm}\left(\hat{\bm{\rho}},\bf{h}\right)$ in Eq.~\eqref{eq:rbmprob} we obtain the following effective Hamiltonian for the visible layer
\begin{equation}
 \begin{split}
H_{\text{rbm}}&\left( \hat{\bm{\rho}}\right)=-\ln\left(\text{Tr}_{{\bf h}}\exp\left(-F_{\text{RBM}}\left(\hat{\bm{\rho}},\bf{h}\right)\right)\right)\\
  =& {\bf B}\cdot\hat{\bm{\rho}} + \sum_{j=1}^{N_h} \ln\left( \sum_{h_j=0}^1 e^{\left(C_j + \sum_{\mu}^M W_{\mu j}\hat{\rho}_{\mu} \right)h_j}\right)\;,
 \end{split}
\end{equation}
where we have implicitly added a (irrelevant) constant energy shift in order to cancel the normalization constant. It is now convenient to express this more compactly as follows
\begin{equation}
H_{\text{rbm}}= {\bf B}\cdot\hat{\bm{\rho}} + \sum_{j=1}^{N_h} K_j^{(2)}\left( \sum_{\mu}^M W_{\mu j}\hat{\rho}_{\mu} \right)\;,
 \label{eq:RBM_E2}
\end{equation}
where in the last line we have defined, similarly to~\cite{Cossu2019}, the cumulant generating function
\begin{equation}
\label{eq:binary_cgf}
\begin{split}
K_j^{(2)}(t) &= \ln\left(\sum_{h=0}^1 e^{\left(C_j+t\right)h}\right)=\ln\left(1+ e^{C_j+t}\right)\\
% &=\sum_{n=1}^\infty \kappa_n^{(2)}\;,
\end{split}
\end{equation}
where the superscript $(2)$ indicates this definition is relevant to binary hidden units only (we will generalize this in Sec.~\ref{sec:categorical}). By performing a Taylor expansion we obtain 
\begin{equation}
\label{eq:binary_cgf_exp}
\begin{split}
&K_j^{(2)}\left(\sum_{\mu=1}^M W_{\mu j}\hat{\rho}_{\mu}\right)=\sum_{n=1}^\infty \frac{\kappa_{jn}^{(2)}}{n!}\left(\sum_{\mu=1}^M W_{\mu j}\hat{\rho}_{\mu}\right)^n \\
&\!\!\!\!=\sum_{n=1}^\infty \frac{\kappa_{jn}^{(2)}}{n!}\!\!\sum_{k_1+\cdots+k_M=n}\!\binom{n}{k_1,\dots,k_M}\!\!\prod_{\mu=1}^M\!\left(W_{\mu j}\hat{\rho}_\mu\right)^{k_\mu}
\end{split}
\end{equation}
where in the second line we used the multinomial expansion and we defined the (binary) cumulants $\kappa_n^{(2)}$ as
\begin{equation}
\kappa_{jn}^{(2)} = \left. \frac{d^n}{dt^n}K_j^{(2)}(t)\right|_{t=0}\;.
\end{equation}
The key observation now is noticing that, due to idempotency of the density operators we have $\hat{\rho}_\mu^n=\hat{\rho}_\mu\;\forall n>0$. The effective interaction coupling strengths of all the interaction terms are then given by the appropriate sums of cumulants, and can be in principle controlled by appropriately choosing the RBM parameters ${\bf B}$,${\bf C}$ and $W$. In~\cite{Cossu2019} the authors derived explicit expressions for up to 3 body interactions in this way.

In this work we use instead a different approach which allow a simpler determination of the induced coupling strengths in the visible layer by solving a linear system of equations obtained by working explicitly in the eigenbasis of the density operators. As we will see more explicitly in the next section, this approach is similar to the one used by Hirsch to find his discrete decomposition for the Hubbard interaction~\cite{Hirsch1983}.

In the next two sections we describe in some detail our approach for the special cases of 2 and 3 body interactions and explain in Appendix~\ref{app:General_EQ} how to construct the mapping for the general case. We comment on possible extension to non-idempotent operators in Appendix~\ref{sec:general_int} but, as we commented in the introduction, this extension is not strictly needed.

\subsection{Two body interactions}
\label{sec:2bd1}

\begin{figure}
 \includegraphics[scale=0.5]{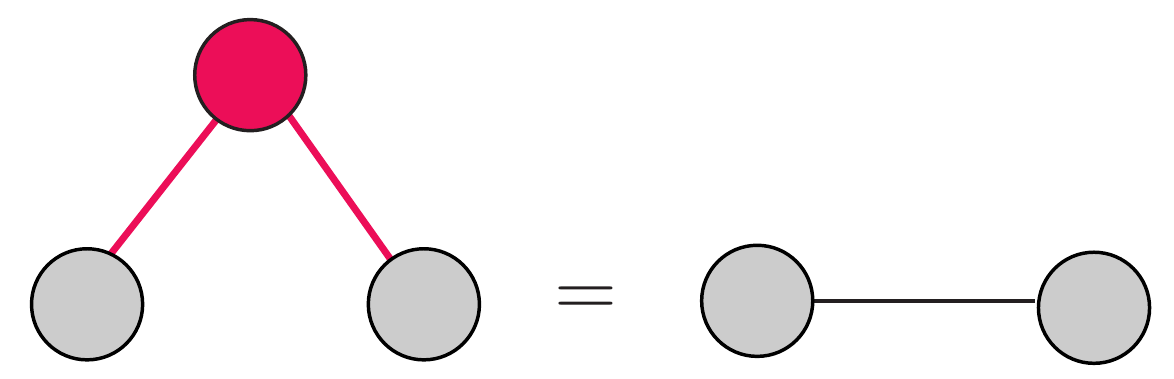}
 \caption{Two-body interaction from the RBM mapping.\label{fig:2bmap}}
\end{figure}

To set the stage we consider now the familiar case of a two body contact interaction, and show how this can be generated by coupling a single binary auxiliary variable (hidden unit) $h\in \{0,1\}$ to a pair of density operators (the visible units) as schematically depicted in Fig.~\ref{fig:2bmap}.  In this case the free energy of the RBM can be written as
\begin{equation}
F^{(2)}_{\text{rbm}}\left(\hat{\rho}_1,\hat{\rho}_2, h\right)=C h + h \sum_{\mu=1}^2 W_{\mu} \hat{\rho}_{\mu}\;,
 \label{eq:H2bd}
\end{equation}
where we neglected possible biases ${\bf B}=\left(B_1,B_2\right)$ on the visible layer since they only contribute to the one-body interaction. If needed, these biases can be used to further control the interactions generated by  Eq.~\eqref{eq:H2bd}. The induced Hamiltonian in the visible layer takes the form
\begin{equation}
\label{eq:htarget}
\begin{split}
H^{(2)}_{\text{rbm}}\left(\hat{\rho}_1,\hat{\rho}_2\right)&=-\ln\left(\sum_{h=0,1}e^{-F^{(2)}_{\text{rbm}}\left(\hat{\rho}_1,\hat{\rho}_2, h\right)}\right)\\
&= A^{(2)} \hat{\rho}_1 \hat{\rho}_2 + A^{(1)}_{1} \hat{\rho}_{1}+A^{(1)}_{2} \hat{\rho}_{2}\;,
\end{split}
\end{equation}
where a direct computation (see also Appendix.~\ref{app:23bd} for additional details) leads to the relations
\begin{equation}
 \begin{split}
 A^{(1)}_{\mu} =& -\ln \left(\frac{e^{-\left(C+W_{\mu}\right)}+1}{e^{-C}+1}\right)\\
 A^{(2)} =& -\ln \left(\frac{e^{-\left(C+\sum_{\mu=1}^2W_{\mu}\right)}+1}{e^{-C}+1}\right)-\sum_{\mu=1}^2 A^{(1)}_{\mu}\;,
 \end{split}
 \label{eq:2bd_general}
\end{equation}
between the $3$ parameters of the RBM and the $3$ coupling constants of the interactions in Eq.~\eqref{eq:htarget}.
These equations can then inverted to determine the RBM parameters needed to produce the wanted two-body term, further details on these derivations are provided in appendix~\ref{app:23bd}.

We note at this point that the discrete Hubbard-Stratonovich transformations from Hirsch~\cite{Hirsch1983} are special cases of Eq.~\eqref{eq:2bd_general} for a particular choice of RBM parameters. For instance, the transformation useful for repulsive interactions (ie. $A^{(2)}>0$) was obtained in~\cite{Hirsch1983} by coupling a classical spin $\sigma=\pm1$ to the spin density $\hat{\rho}_{sp}=\hat{\rho}_\uparrow-\hat{\rho}_\downarrow$, the resulting partition function is
\begin{equation}
Z_{sp} =\frac{1}{2} \sum_{\sigma=\pm}\exp\left(2a\sigma\hat{\rho}_{sp}\right)=\cosh\left(2a\hat{\rho}_{sp}\right)
\end{equation}
where the coupling constant $a$ is given by 
\begin{equation}
\label{eq:a2}
 \tanh(a)^2 = \tanh\left(\frac{A^{(2)}}{4}\right)\;,
\end{equation}
and $Z_{sp}$ contains additional one body terms corresponding to $A^{(1)}_1=A^{(1)}_2=-A^{(2)}/2$. Using the RBM model of Eq.~\eqref{eq:H2bd} a similar transformation, with different additional one-body terms $A^{(1)}_1-2a=A^{(1)}_2+2a=-A^{(2)}/2$, can be obtained choosing $C=0$ and $W_1=-W_2=4a$. %A similar correspondence can be found for the transformation obtained by coupling the classical coupled to the total spin instead which is useful for attractive interactions (ie. $A^{(2)}<0$). 

Note that there is a continuous set of RBM parameters which will lead to the same two-body coupling $A^{(2)}$ and different induced one-body terms in Eq.~\eqref{eq:2bd_general}. These possibly unwanted one-body contributions to the visible layer Hamiltonian can then be removed by adding the appropriate bias terms ${\bf B}$ in Eq.~\eqref{eq:H2bd}. 

Before moving on to the results for the case of three-body forces, we want to show how the RBM mapping is not limited to contact interactions but can be immediately applied to situations where the operators associated with the visible layer are either different idempotent operators (for instance projection operators) or, equivalently, involutive operators for which $\hat{O}^2=\mathbb{1}$. To the latter class belong for instance the Pauli operators (and more generally any element of the Pauli group formed by arbitrary tensor products of Pauli matrices) which are routinely used to model the spin-isospin structure of nuclear interactions. For instance, consider the case of the following repulsive interaction acting on a pair of spins
\begin{equation}
U\left(\vec{\sigma}_1\cdot\vec{\sigma}_2\right) = U\sum_{d=1}^3 {\sigma}_d\otimes{\sigma}_d\;.
\end{equation}
The propagator associated with it can now be written using 3 binary auxiliary variables,
\begin{equation}
 \begin{split}
  e^{-U \alpha_t \left( \vec{\sigma}_1\cdot\vec{\sigma}_2\right)}=&\frac{e^{-3 U \alpha_t}}{8}\prod_{d=1}^3\sum_{h_d=0}^1 e^{a (2h_d-1)({\sigma}_d\otimes \mathbb{1}- \mathbb{1} \otimes \sigma_d)}.
 \end{split}
 \label{eq:2bdpauli}
\end{equation}
In the expression above $\alpha_t$ is the imaginary-time step size used in the calculation, $\mathbb{1}$ is the identity matrix, and $a$ is given by Eq.~\eqref{eq:a2} with $A^{(2)}=4U \alpha_t$. %In Sec.~\ref{sec:general_int} we show how such mappings can be generalized.

As we have seen, the complete generality of the mapping Eq.~\eqref{eq:H2bd} allows its application to simplify a large number of pair-interacting many-body theories.

\subsection{Three Body interactions}
\label{sec:3bd}

The presence of three-body contact interactions is of fundamental importance in low-energy effective theories of nuclear physics and some bosonic cold gases in order to avoid the Thomas collapse of finite clusters~\cite{Thomas1935} and ensure renormalizability of the Hamiltonian~\cite{Hammer2013,Carlson2017,vanKolck2019}.
Due to the inability to use the standard Hubbard-Stratonovich or Hirsch's transformations, the inclusion of these 3 body contact interactions in auxiliary field calculations is not straightforward. Here we will briefly review the approach used in Lattice EFT calculations~\cite{Lee2009} as a relevant example. In order to integrate out the fermionic degrees of freedom and carry out the auxiliary field calculation one needs to decouple the interacting part of the partition function
\begin{equation}
\label{eq:zint}
Z_{int} = \exp\left(- \frac{U\alpha_t}{2} \sum_{a,b}\hat{\rho}_a \hat{\rho}_b- \frac{V\alpha_t}{6} \sum_{a,b,c}\hat{\rho}_a \hat{\rho}_b\hat{\rho}_c\right),
\end{equation}
which contains both a two-body and a three-body interaction, in terms of one-body fermionic operators only. Similarly to the previous section,  $\alpha_t$ is the size of the imaginary-time step.

\begin{figure}%[ht]
 \includegraphics[scale=0.5]{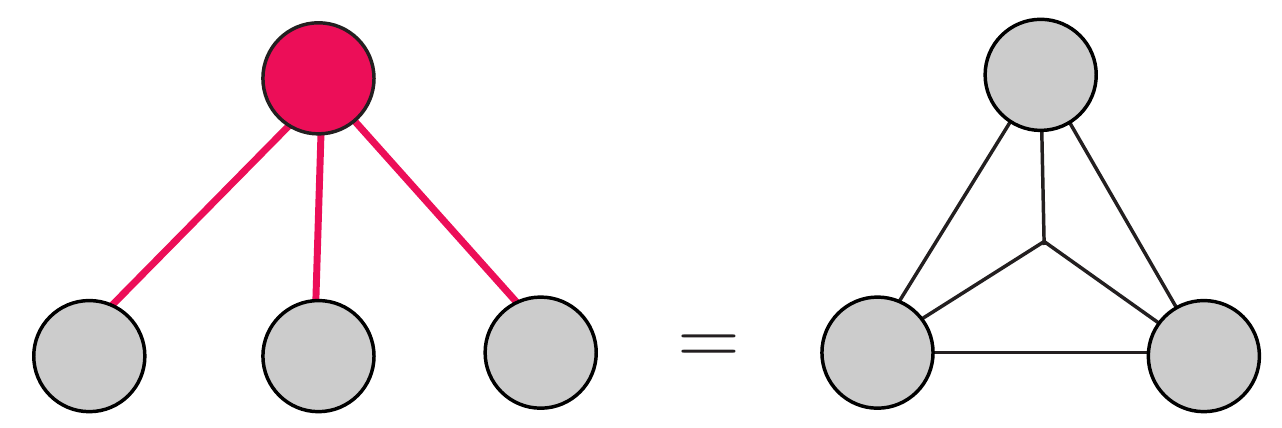}
 \caption{Three-body interaction from the RBM mapping.\label{fig:3bmap}}
\end{figure}   

The strategy proposed in~\cite{Chen2004} is to represent $Z_{int}$ using a single continuous auxiliary field
\begin{equation}
  Z_{int}= \int_{-\infty}^{\infty}dh P(h) \exp\left(h\sum_a\hat{\rho}_a\right)\;,
\end{equation}
for an appropriately chosen probability distribution function $P(h)$ of the auxiliary field. In~\cite{Chen2004} it was shown how one can use the solution of the truncated Hamburger moment problem to find the wanted distribution. In order for this representation to exist one needs the interaction couplings in Eq.~\eqref{eq:zint} to satisfy
\begin{equation}
\label{eq:3bcond}
V^2<-2 \alpha_t U^3\;,
\end{equation}
which then prevents this to work for repulsive two body couplings $U\geq0$. In the more common case of an attractive two body interaction, this mapping can be exploited only for a weak enough three body coupling $V$ and large enough imaginary-time step $\alpha_t$. Notably, for other classes of QMC methods based on auxiliary fields such as AFDMC (see eg.~\cite{Carlson2015}) this is not possible in general due to the inability to introduce explicitly an energy cutoff as is done in a lattice formulation. We note in passing that an alternative auxiliary field representation of the partition function of Eq.~\eqref{eq:zint} can be obtained using the DBM mapping introduced in ~\cite{Yoshioka2019}. Also in this case the mapping can be performed only for weak enough three-body interaction: $|V|<-3U$ and attractive two body coupling $U<0$.

Thanks to the flexibility of the RBM mapping scheme we can instead treat the two and three body interactions separately using different hidden variables. In the rest of this section we focus then directly on a mapping that reduces the three-body term by using a single binary unit coupled to three density operators as depicted in Fig.~\ref{fig:3bd_categorical}. The resulting free energy of the RBM is similar to Eq.~\eqref{eq:H2bd} and reads
\begin{equation}
F^{(3)}_{\text{rbm}}(\hat{\rho}_1,\hat{\rho}_2,\hat{\rho}_3, h)=C h + h \sum_{\mu=1}^3 W_{\mu} \hat{\rho}_{\mu}\;,
\label{eq:hrbm3}
\end{equation}
while the target Hamiltonian can be written as
\begin{equation}
 \begin{split}
\!\!H^{(3)}_{\text{rbm}}(\hat{\bm{\rho}})=& A^{(3)}\hat{\rho}_1 \hat{\rho}_2 \hat{\rho}_3+\!\!\sum_{\mu<\nu}^{3} A^{(2)}_{\mu \nu} \hat{\rho}_{\mu} \hat{\rho}_{\nu} + \!\!\sum_{\mu=1}^{3} A^{(1)}_{\mu} \hat{\rho}_{\mu}.
  \end{split}
  \label{eq:H3bd}
\end{equation}
Since we are only interested in producing a target three-body term $A^{(3)}$ while removing the unwanted interactions using lower order mappings, it is convenient to simplify this to
\begin{equation}
\!\!H^{(3)}_{\text{rbm}}(\hat{\bm{\rho}})= A^{(3)}\hat{\rho}_1 \hat{\rho}_2 \hat{\rho}_3+A^{(2)}\!\sum_{\mu<\nu}^{3} \hat{\rho}_{\mu} \hat{\rho}_{\nu} + A^{(1)}\!\sum_{\mu=1}^{3} \hat{\rho}_{\mu},
\end{equation}
with mode-independent one and two body couplings. In the same way we set $W_\mu=W$ in Eq.~\eqref{eq:hrbm3}.
As we did for the two-body interactions above, a direct calculation leads to the following relations between the physical couplings and the RBM parameters (see also Appendix~\ref{app:General_EQ} for the general case):
\begin{equation}
 \begin{split}
  A^{(1)}=&-\ln \left(\frac{e^{-\left(C+W\right)}+1}{e^{-C}+1}\right)\\
  A^{(2)}=&-\ln \left(\frac{e^{-\left(C+2W\right)}+1}{e^{-C}+1}\right) - 2 A^{(1)}\\
  A^{(3)}=&-\ln \left(\frac{e^{-\left(C+3W\right)}+1}{e^{-C}+1}\right)-3 A^{(2)}- 3 A^{(1)}.
 \end{split}
 \label{eq:3bd_symmetric}
\end{equation}
As before, in order to find the needed RBM parameters these relations need to be inverted and, depending on the sign of the physical three body coupling, one needs to make appropriate
choices for the range of the RBM parameters. For instance, if we are interested in an attractive interaction, $A^{(3)}<0$, we can set  $C=-2W$ while in the case of a repulsive interaction,  $A^{(3)}>0$, we can take $C=-W$. In both cases we obtain the same magnitude for the three body coupling, $|A^{(3)}|=\ln \left(\cosh ^4\left(\frac{W }{2}\right) \text{sech}(W )\right)$.

The two body interaction $A^{(2)}$ can be removed by expressing it in terms of additional auxiliary variables using the identities in Eq.~\eqref{eq:2bd_general}. Note that with this technique no constraint like Eq.~\eqref{eq:3bcond} needs to be imposed on the value of the physical couplings. More details on these derivations, and their extension to the case of categorical hidden units, are provided in appendix~\ref{app:23bd}. 

We now turn to the discussion of categorical variables and provide a justification for their use in practical applications.

\subsection{Categorical hidden variables}
\label{sec:categorical}

We proceed now to generalize the RBM mappings derived above for hidden binary variables to the more general case of  categorical hidden variables which take values on a larger range $\{0,1,2,...,\mathcal{K}-1\}$. From the form of the general RBM free energy Eq.~\eqref{eq:RBM_E1} we can expect that the effect of increasing the magnitude of the hidden variable, controlled by $\mathcal{K}$, is to correspondingly increase the energy term proportional to $\bf{C}$ and $W$. This, in turn, will increase the magnitude of the induced physical coupling at fixed RBM parameters. We expect this property to be useful for instance in the sampling process as it allows for smaller energy gaps when updating the hidden layer. Another instance when this could possibly be useful is to minimize the systematic error introduced by Trotter like decompositions of the imaginary-time propagator (see eg.~\cite{Carlson2015}) but we leave a more detailed exploration of these possibilities to future work.

\begin{figure}
  \includegraphics[scale=0.4]{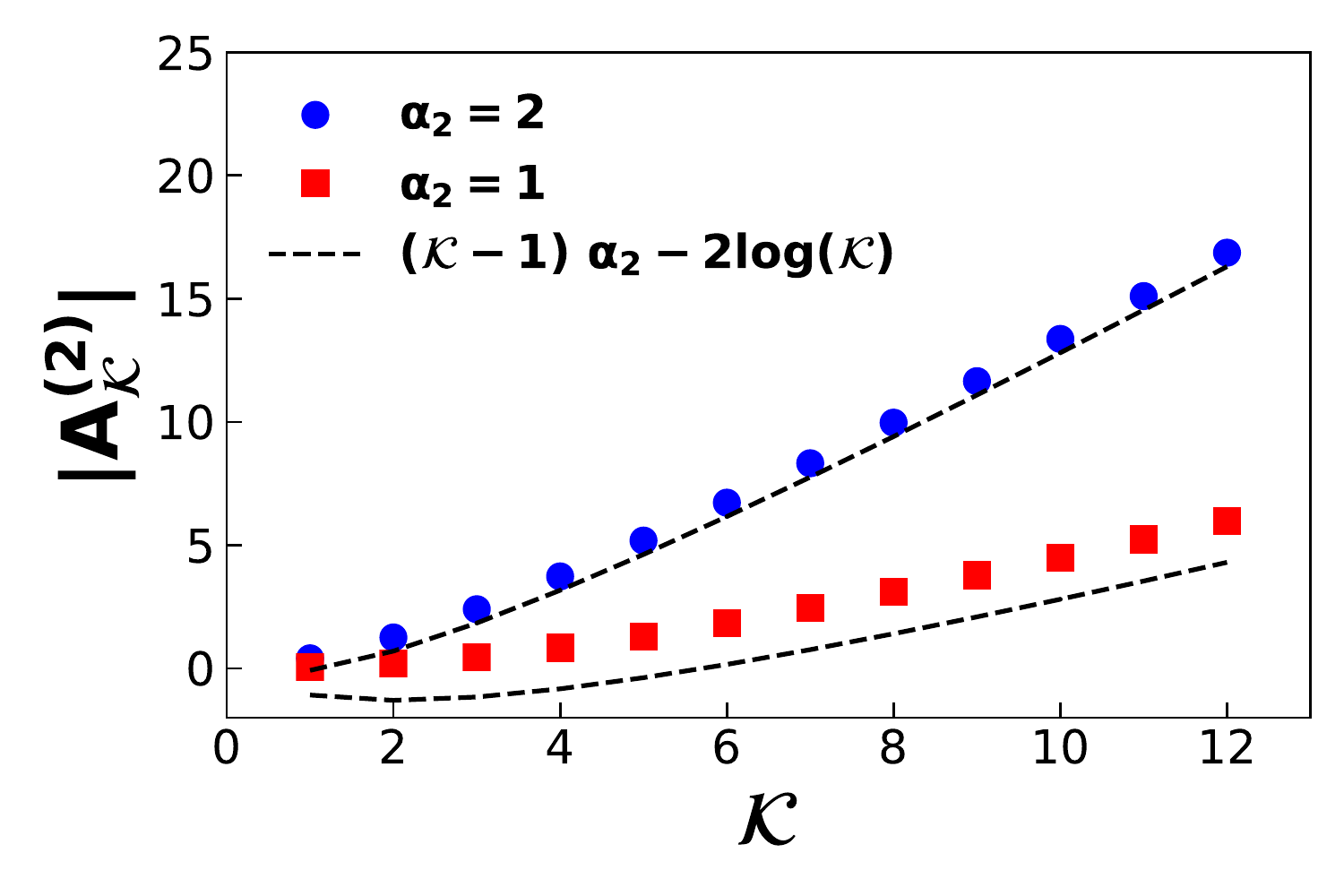}
 \caption{Two body coupling as function of $\mathcal{K}$. The dashed line corresponds to the lower bound in Eq.~\eqref{eq:2blbound}.}
 \label{fig:2bd_categorical}
\end{figure}

The modified expression for the induced one and two-body coupling when coupling a pair of visible units with a single categorical hidden variable, corresponding to the generalization of Eq.~\eqref{eq:2bd_general}, can be compactly written as 
\begin{equation}
\label{eq:2bmap_cat_one}
 \begin{split}
  A^{(1)}_{\mu}=&-\ln \left(\frac{e^{-\mathcal{K} (C+W_{\mu})}-1}{e^{-\left(C+W_{\mu}\right)}-1}\right)\\
  &+\ln \left(\frac{e^{-\mathcal{K}C}-1}{e^{-C}-1}\right)\;.
 \end{split}
\end{equation}
for the one-body term, while for the two-body we have
\begin{equation}
\label{eq:2bmap_cat_two}
 \begin{split}
  A_{\mathcal{K}}^{(2)}=&-\ln \left(\frac{e^{-\mathcal{K} \left(C+\sum_{\mu=1}^2W_{\mu}\right)}-1}{e^{-\left(C+\sum_{\mu=1}^2W_{\mu}\right)}-1}\right)\\
  &+\ln \left(\frac{e^{-\mathcal{K}C}-1}{e^{-C}-1}\right)-\sum_{\mu=1}^2A^{(1)}_{\mu}\;.
 \end{split}
\end{equation}
Note that, whenever the argument of the exponentials in Eq.~\eqref{eq:2bmap_cat_one} and Eq.~\eqref{eq:2bmap_cat_two} becomes 0, one need to take the limit continuously. To show the effect of increasing the range of the hidden variable on the induced coupling constant, we plot in Fig.~\ref{fig:2bd_categorical} the magnitude of the two body coupling as a function of $\mathcal{K}$ for two one parameter families of RBMs:
\begin{itemize}
 \item for attractive interactions $A^{(2)}_\mathcal{K}<0$ we take
\begin{equation}
\label{eq:onebpar_a}
W_1=W_2=-C=\alpha_2>0\;,
\end{equation}
 \item for repulsive interactions $A^{(2)}_\mathcal{K}>0$ we take
\begin{equation}
\label{eq:onebpar_r}
W_1=-W_2=\alpha_2>0\quad C=0\;.
\end{equation}
\end{itemize}
Note that these choices were motivated only by the need to obtain all possible two-body coupling by tuning a single parameter $\alpha_2$ so that we could present the results compactly in Fig.~\ref{fig:2bd_categorical}. This is therefore a completely general construction and it is possible that similar results could be obtained by performing different choices. 

As the results show, the magnitude $\left|A_{\mathcal{K}}^{(2)}\right|$ increases almost linearly with $\mathcal{K}$ for a fixed choice of the RBM parameter $\alpha_2$. As an easier proxy to understand the expected behavior, and especially the apparent asymptotically linear growth with $\mathcal{K}$, we also show in Fig.~\ref{fig:2bd_categorical} the following lower-bound
\begin{equation}
\label{eq:2blbound}
\left|A_{\mathcal{K}}^{(2)}\right| \geq\left( \mathcal{K} -1\right) \alpha_2 - 2 \ln\left(\mathcal{K}\right)\;,
\end{equation}
valid for both parametrizations Eq.~\eqref{eq:onebpar_a} and Eq.~\eqref{eq:onebpar_r}. Indeed we find the approximately linear increase with $\mathcal{K}$ together with a mild logarithmic correction. A complete proof of this lower-bound, and the one for three-body interactions, is provided Appendix~\ref{app:categorical}.

\begin{figure}%[ht]
  \includegraphics[scale=0.4]{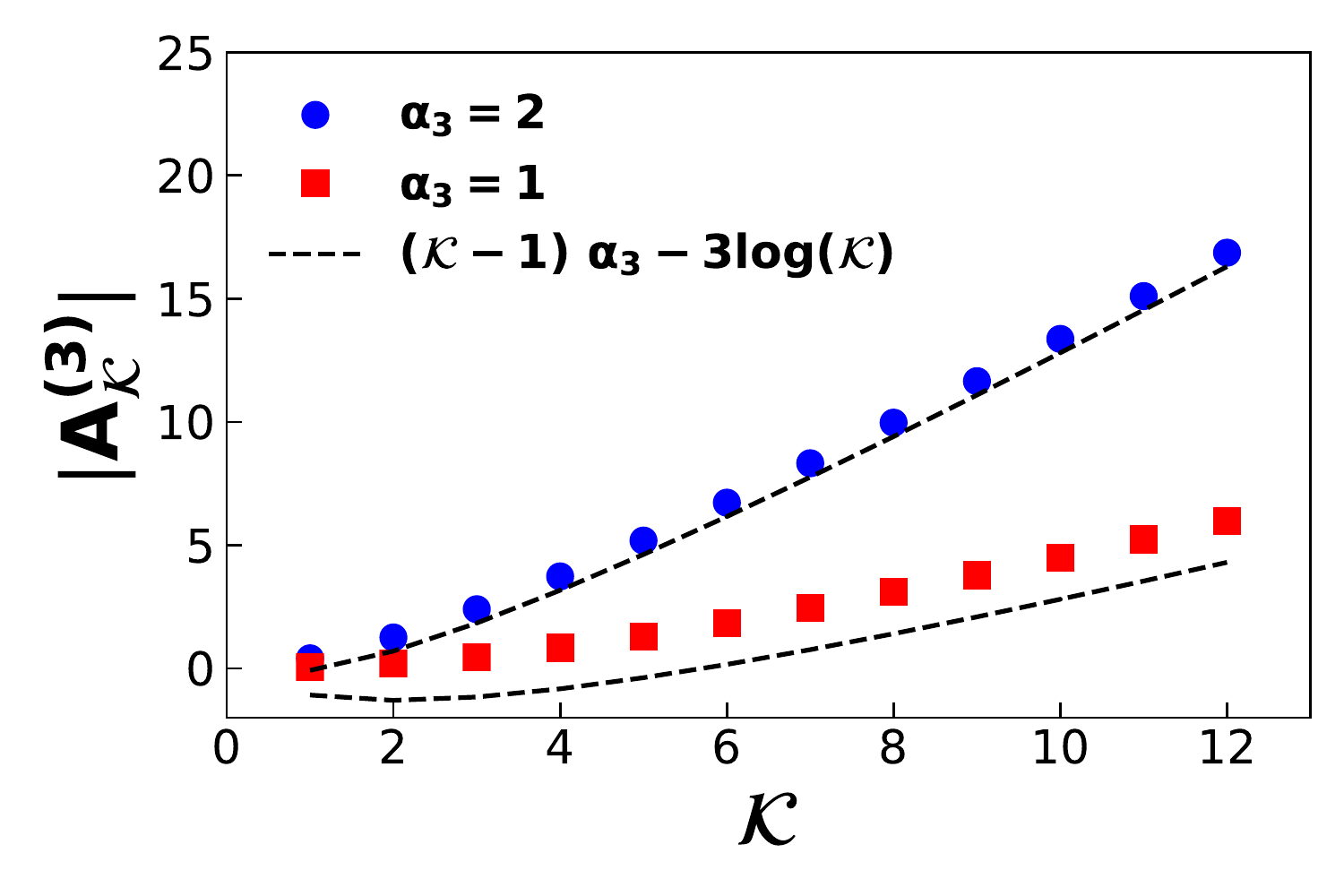}
 \caption{Three body coupling as a function of $\mathcal{K}$. The dashed line corresponds to the lower bound in Eq.~\eqref{eq:3blbound}.}
 \label{fig:3bd_categorical}
\end{figure}

Next, we turn our attention to the more complicated case of the three body interaction induced by a single categorical variable. The mapping between physical coupling constants and RBM parameters remains the same as Eq.~\eqref{eq:2bmap_cat_one} at the one-body level while the two and three body couplings become (see also Appendix~\ref{app:23bd} for a full derivation) 
\begin{equation}
\label{eq:3bmap_cat}
 \begin{split}
  A^{(2)}_{\mu \nu}=&-\ln \left(\frac{e^{-\mathcal{K}\left(C+W_{\mu}+W_{\nu}\right)}-1}{e^{-\left(C+W_{\mu}+W_{\nu}\right)}-1}\right)\\
  &+\ln\left(\frac{e^{-\mathcal{K}C}-1}{e^{-C}-1}\right) - A_{\mu} - A_{\nu}\\
   A_\mathcal{K}^{(3)}=&-\ln\left(\frac{e^{-\mathcal{K} \left(C+W_1+W_2+W_3\right)}-1}{e^{-\left(C+W_1+W_2+W_3\right)}-1}\right)\\
  +&\ln \left(\frac{e^{-\mathcal{K}C}-1}{e^{-C}-1}\right) -\sum_{\mu=1}^{3}\sum_{\nu > \mu}^3  A_{\mu \nu}-\sum_{\mu=1}^3A_{\mu}\;,
 \end{split}
\end{equation}
and similarly to above the induced physical coupling grows approximately linearly with $\mathcal{K}$. This is shown in Fig.~\ref{fig:3bd_categorical} where we plot  both the actual magnitude $\left|A_{\mathcal{K}}^{(3)}\right|$ and the lower-bound
\begin{equation}
\label{eq:3blbound}
\left|A_{\mathcal{K}}^{(3)}\right| \geq\left( \mathcal{K} -1\right)\alpha_3 - 3 \ln\left(\mathcal{K}\right)\;.
\end{equation}
for two values of the RBM parameter, $\alpha_3=1,2$. In the three body case we've also chosen a simple one-parameter family of RBMs given by the choices:
\begin{itemize}
 \item for attractive interactions $A^{(3)}_\mathcal{K}<0$ we take
\begin{equation}
W_1=W_2=W_3=\alpha_3>0\quad C=-2\alpha_3
\end{equation}
 \item for repulsive interactions $A^{(3)}_\mathcal{K}>0$ we take
\begin{equation}
W_1=W_2=W_3=\alpha_3>0\quad C=-\alpha_3\;.
\end{equation}
\end{itemize}
As for the case of the two-body term in Fig.~\ref{fig:3bd_categorical}, we see that the lower-bound is relatively tight. A complete derivation of Eq.~\eqref{eq:3blbound} is also provided in Appendix~\ref{app:categorical}.

% In fig.~\ref{fig:3bd_categorical} we plot the magnitude of three body coupling as function of $\mathcal{K}$ and the lower bound value from appendix~\ref{app:categorical}
% As both plots show, we expect an almost linear decrease of the RBM parameter as function of $\mathcal{K}$ for the same physical coupling.
% As already mentioned, in practice, increasing $\mathcal{K}$ allows us to implement small RBM parameters for large two and three body forces, which should aid Monte Carlo sampling and keep systemic statistical uncertaintes under control.

Similarly to the results described in the previous two sections (see Eq~\eqref{eq:2bdpauli}), this more general construction can be easily extended to many-body operators composed by involutive operators with minimal changes.

Before concluding this section, we want to also point out that it is relatively straightforward to use the expansion of the cumulant generating function to get the effective Hamiltonian for the visible layer as proposed in~\cite{Cossu2019} and explained in detail at the beginning of Sec.~\eqref{sec:conjecture}. Since this can be of more general interest, we provide here the necessary generalization to the cumulant generating function of Eq.~\ref{eq:binary_cgf} which for categorical variables becomes
\begin{equation}
\label{eq:categorical_cgf}
\begin{split}
K^{(\mathcal{K})}_j(t) &= \ln\left(\sum_{h=1}^{\mathcal{K}-1}\exp\left(-h(C_j+t)\right)\right)\\
&= \ln\left(\frac{1-\exp\left(-\mathcal{K}(C_j+t)\right)}{1-\exp\left(-(C_j+t)\right)}\right)\;.
\end{split}
\end{equation}
After this change the polynomial expansion from Eq.~\eqref{eq:binary_cgf_exp} remains essentially equivalent, with the only difference that the value of the cumulants $\kappa_{jn}^{(\mathcal{K})}$ will now need to be extracted by taking appropriate derivatives of $K^{(\mathcal{K})}_j(t)$.

\section{Many body forces implementation on Quantum Annealers}
\label{sec:qa}

Quantum annealing is an algorithm for finding approximate solutions to non convex optimization problems using a quantum generalization of classical simulated annealing~\cite{Finnila1994,Kadowaki1998,farhi2000}. 
This approach has been applied in the past to hard optimization problems ranging from computing Ramsey numbers~\cite{Gaitan2012} to finding low energy configurations of proteins~\cite{Ortiz2012}. Used as an analog simulator, a quantum annealer was also recently used to study phase transitions in lattice models~\cite{King2018}.

The underlying idea of quantum annealing based optimization is to first encode the solution of the problem at hand in the ground state of a k-local Hamiltonian acting on spins. This Hamiltonian is then simulated by a physical system that is cooled down to reach a low energy state which provides a good approximate solution to the optimization problem. In order to decrease the time required to find a solution of this approach, the physical system is initialized as the ground state of an auxiliary Hamiltonian $H_A$ and then adiabatically evolved to the ground state of the problem Hamiltonian $H_P$. If this is done slowly enough to avoid non-adiabatic transitions, the final state will be the ground state of $H_P$ corresponding to the optimal solution~\cite{Farhi2001}.
In some situations fast non-adiabatic transitions can also be used to accelerate the convergence to a close approximation to the final ground state~\cite{Das2005,Chandra2010}.

Here we will focus more specifically on the class of quantum annealers that uses a tunable transverse field Ising model to implement this idea. An example of this is the device manufactured by D-Wave Systems~\cite{Johnson2011}. The time-dependent spin Hamiltonian is given by
\begin{equation}
 \begin{split}
  \mathcal{H}(\tau)=&B_x(\tau)H_A+B_z(\tau)H_P\\
  H_A=&-\sum_i\sigma_i^x\\
  H_P=&\sum_{ij}J_{ij}\sigma_{i}^z \sigma_j^z + \sum_i h_i \sigma_i^z
 \end{split}
 \label{eq:HAB}
\end{equation}
where $\sigma_{i}^{x,z}$ are Pauli matrices that operate on spin or
qubit $i$. Note that the optimization problem is encoded in the diagonal Hamiltonian $H_P$ by tuning appropriately both the pairwise couplings $J_{ij}$ as well as the on-site fields $h_i$. The auxiliary Hamiltonian $H_A$ is chosen for two reasons, it's ground state is a complete superposition of all the basis states and it is easy to prepare by applying a physical transverse field. The time evolution is then performed by modifying the applied magnetic fields $B_x$ and $B_z$ starting from $B_x(0)\ll B_z(0)$ and evolving to $B_x(1)\gg B_z(1)$. Here $\tau = t/t_a$ , where
$t$ is the physical time and $t_a$ is the annealing time:
the time that it takes to perform the transition from the Hamiltonian $H_A$ into the problem Hamiltonian $H_P$. 

Due to the presence of two-body spin coupling in $H_P$, this system can be employed to approximately solve Quadratic unconstrained binary optimization (QUBO) problems. The goal of this section is to show how the RBM mapping we derived in this work can be used to increase the representation power of the problem Hamiltonian $H_P$ by implementing higher order diagonal interactions using some of the available qubits as auxiliary spins in the hidden layer. This process of reducing many-body interactions to two-body terms has been extensively explored in the past (see eg.~\cite{Perdomo2008,Bian2014,Chancellor2017} and~\cite{Dattani2019} for a recent summary). Known mappings are limited in their application to specific signs of the many-body coupling, for instance only interactions with positive cubic terms are known to embed into the Chimera topology without using additional auxiliary qubits for the embedding~\cite{Dattani2019}. Here we show that our RBM construction can be applied at the same cost for cubic terms of both signs. The technique can also be easily extended to higher-body interactions.

For illustration purposes, we will consider a simple extension to the QUBO model to contain also cubic terms implemented as effective couplings among three spins. In particular consider the following cost function
\begin{equation}
 \begin{split}
  E=a \sum_i^3  q_i + b \sum_{i<j}^3 q_i q_j + c\  q_1 q_2 q_3\;,
 \end{split}
\end{equation}
with only 3 binary variables $q_i$. This can be converted into a generalized classical Ising model by introducing the spin variables $s_i=2q_i-1$, and then mapped to a diagonal quantum Hamiltonian acting on physical spins ($s_i \rightarrow \sigma^z_i$):
\begin{equation}
 \begin{split}
  H=&A \sum_i^3 \sigma^z_i + B \sum_{i<j}^3 \sigma^z_i \sigma^z_j + C \sigma^z_1 \sigma^z_2 \sigma^z_3,\\
  A=&2(a+b-c), \ B=4(b-c), \ C=8c\;.
 \end{split}
 \label{eq:Hsising}
\end{equation}
Using the notation from Ref.~\cite{Dattani2019}, this Hamiltonian corresponds to the $K_4$ gadget.

\begin{figure}[ht]
 \includegraphics[scale=0.5]{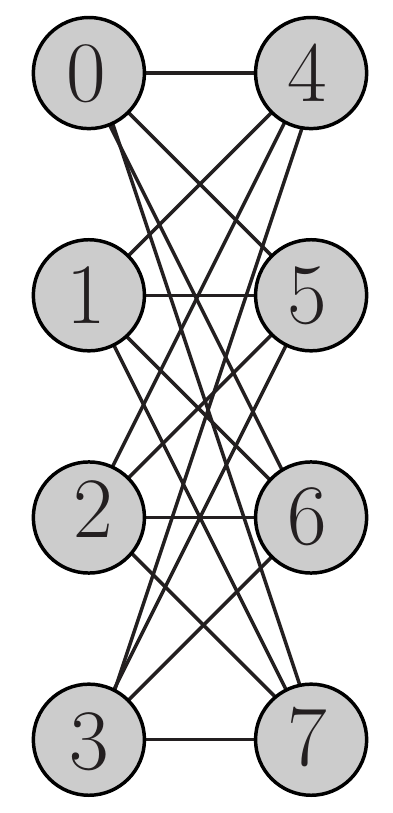}
 \caption{Cell of chimera graph in D-Wave systems.}
 \label{fig:chimera_cell}
\end{figure}

Due to hardware restrictions the topology of the interactions between qubits is restricted to only a finite set. In this exploratory work we consider the Chimera graph~\cite{boothby2016} found on D-Wave devices, the scheme can however be generalized to different topologies.
In the Chimera graph topology, all qubits in one partition are linked to the qubits in the other partition but not among themselves as shown in fig.~(\ref{fig:chimera_cell}).

\begin{figure}[ht]
 \includegraphics[scale=0.5]{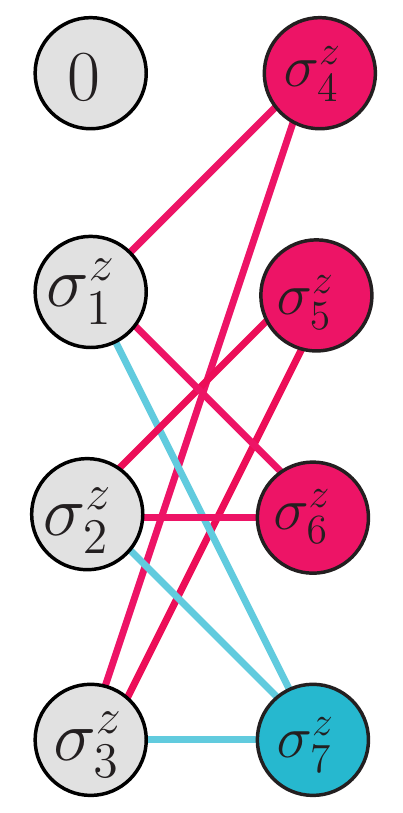}
 \caption{Three body interactions implemented on the chimera topology.}
 \label{fig:chimera_3bd}
\end{figure}

Given the coupling map in Fig.~\ref{fig:chimera_cell}, we can implement the $K_4$ gadget using 4 ancilla qubits:
\begin{itemize}
    \item qubits 1,2,3 represent the logical qubits
    \item qubit 4 mediates the pair interaction between $(1,3)$
    \item qubit 5 mediates the pair interaction between $(2,3)$
    \item qubit 6 mediates the pair interaction between $(1,2)$
    \item qubit 7 mediates the cubic interaction
\end{itemize}
This is also depicted in Fig.~\ref{fig:chimera_3bd}. The full Hamiltonian takes the following form
\begin{equation}
\begin{split}
H_P &= W_1 \sum_{i=1}^3\sigma^z_i -\text{sgn}(C)W_3\sigma^z_7\\
& + \left|W_2\right| (\sigma^z_1 + \text{sgn}(A_3) \sigma^z_2)\sigma^z_4\\
& + \left|W_2\right| (\sigma^z_2 + \text{sgn}(A_3) \sigma^z_3)\sigma^z_5\\
& + \left|W_2\right| (\sigma^z_1 + \text{sgn}(A_2) \sigma^z_2)\sigma^z_6\\
& + W_3 \sum_{i=1}^3\sigma^z_i\sigma^z_7\\
\end{split}
\end{equation}
where the first line represent on-site energy shifts, the second through fourth represent the pair interactions induced by qubits $4-6$ and the last one contains the pair interactions with qubit 7 responsible for the three-body term. The coupling terms can then be obtained using the RBM mapping described above (see also Appendix~\ref{app:General_EQ}). In practice we first compute the coefficient $W_3$ from the relation
\begin{equation}
  |C|= \frac{1}{8} \ln \left(\cosh ^4(2 W_3) \text{sech}(4 W_3)\right)\;,
\end{equation}
and the other parameters can then be found using
\begin{equation}
 \begin{split}
  A_2=&\frac{\text{sgn}(C)}{4} \ln (\cosh (4 W_3))-2 B,\\
  W_2=&\frac{\text{sgn}(A_2)}{2} \ln \left(\cosh \left(A_2\right)\right),\\
  W_1=&-\frac{\text{sgn}(C)}{8} \ln (\cosh (4 W_3))-A.
 \end{split}
 \label{eq:bd3_id2}
\end{equation}
This mapping was derived from eqs.~(\ref{eq:2bd_generalv2}) and (\ref{eq:3bd_generalv2}) by replacing $0$ with $-1$.
If we trace out the auxiliary qubits belonging to the hidden layer we recover the effective Hamiltonian Eq.~\eqref{eq:Hsising}. Furthermore, one could reduce the ancilla requirement even further by using a single ancilla to embed direct pair interactions between qubit 1 and qubits 2 and 3 while using a single ancilla to represent the missing two-body term between qubit 2 and qubit 3 (qubit 5 in the mapping above). This would have the same qubit overhead as the best method from Ref.~\cite{Dattani2019} while requiring only one ferromagnetic coupling for the embedding.

Note that the RBM identities are valid for any direct product of Pauli matrices $\sigma^{x,y,z}$ as we discussed in the introduction. This means that extensions like the one presented here for the simple Hamiltonian Eq.~\eqref{eq:HAB} can be generalized to the interesting situation where non stoquastic physical interactions are available.

In practical applications an effective temperature $T_{\text{eff}}$ is usually introduced to properly describe the generated statistical distribution~\cite{Amin2015,Ortiz2016}. Since the effective Hamiltonian is obtained by tracing over the global partition function including the auxiliary spins, one will need to estimate $T_{\text{eff}}$ first (see eg.~\cite{Benedetti2016}) in order to appropriately tune the RBM parameters. We leave the exploration of these issues for future work.

Before concluding we note that the RBM mapping derived in this paper can also be used to show the equivalence between various gadgets presented in Ref.~\cite{Dattani2019}. For instance, it is simple to use the RBM mapping to show that the gadget $K_6-e$ is equivalent, upon appropriate rescaling of the interactions, to the simpler gadget $K_5$.
\section{On training the RBM for physical systems}
\label{sec:train}
 In this final section we turn our attention to machine learning, specifically on training neural architectures for physical systems with many-body interactions. The relative entropy is commonly used in the literature as an objective function which measures the discrepancy between the target data distribution and the approximation given by the neural architecture; the smaller the discrepancy, the smaller its value. For energy based models like the RBM, the probability of the model depends on the accurate evaluation of the partition function in Eq.~\eqref{eq:zbeta}, which is intractable as it requires all possible data samples. The inability to have a reliable estimate of the relative entropy, makes it impossible to get its gradients with respect to the
 RBM parameters. These gradients are needed to update the parameters towards optimal values. 
In practice, approximate gradients are obtained through contrastive divergence (CD)~\cite{Hinton2002}. This procedure was employed in~\cite{Cossu2019} to learn the Ising model from samples generated with the Swendsen-Wang algorithm. However, CD might not always be a good approximation~\cite{Bengio2009,Sutskever2010} and the lack of a point estimator for an objective function makes it harder to asses whether training has converged. In addition, in many cases of interest, Monte Carlo sampling can be highly autocorrelated, so we might not be able to find algorithms that provide samples which reliably represent the physical distribution.

In this section we provide several suggestions for how to improve training. We can use our knowledge of the original Hamiltonian to design a network with only a small number of parameters, independent of system size. As a first step, borrowing from the results presented in previous sections, we construct a sparse RBM whose architecture is based on the interaction order in the physical Hamiltonian. For each interaction in the physical system there is a corresponding auxiliary variable. This architecture is compared to an RBM with the number of auxiliary variables equal to the number of visible variables with all--to--all connection between the two layers.

We then devise an objective function that does not depend on the partition function and is inspired by Monte Carlo acceptance and rejection method. As we demonstrate through experiments, this objective function makes training much easier as it makes use of our understanding of the physical system. Furthermore, since it can be easily evaluated during training, this objective function can serve as an indicator for convergence. 

Lastly, we explore the scenario when we can not rely on Monte Carlo samples collected for training and show that learning is still possible.

As an illustration, we consider a classical anisotropic 2D Ising model in the eigenbasis of $\sigma^z|s\rangle=s|s\rangle$,
\begin{equation}
 \begin{split}
  Z_{2D}=&\mathcal{N}\exp\left(\sum_{i=1}^{L_X} \sum_{j=1}^{L_T} \Lambda_X s_{i,j} s_{i+1,j} +\Lambda_T s_{i,j} s_{i,j+1}\right),\\
 \end{split}
\end{equation}
This partition function approximates the 1D Ising model with a transverse field through Trotter expansion~\cite{Suzuki1976},
 \begin{equation}
 \begin{split}
  H_{1D}=&-J\sum_{i}\sigma^z_i \sigma^z_{i+1} - B \sum_i \sigma^{x}_i,\\
  \Lambda_X =&\frac{\beta}{L_T} J, \
  \Lambda_T = \frac{1}{2}\ln \left( \coth\left(\frac{\beta}{L_T} B \right) \right).
 \end{split}
 \label{eq:1DT}
\end{equation}
The  Hamiltonian of this system is a simpler version of Eq.~\eqref{eq:HAB}.
$L_X$ is the length of the Ising chain and $L_T$ is the number of Trotter steps and is the second dimension in the classical model. From appendix~\ref{app:23bd} we can derive the mapping between the physical couplings $\Lambda_{X,T}$ and the respective RBM parameters. 

To verify the RBM learning procedure we compare the relative error in reproducing these couplings,  $\delta \Lambda = \left|\frac{\Lambda-\Lambda_{\text{RBM}}}{\Lambda}\right|$. The results from~\cite{Cossu2019} show that CD training can have a good performance for the Ising model, but this is not guaranteed in more complicated settings.

Following the procedure outlined in~\cite{Cossu2019}, we generate data of $10^5$ samples with $L_X=L_T=28$ and $J=B=2\beta=1$ through local Monte Carlo sampling and train in batches of 500 samples with a learning rate of $0.001$ using CD with 10 Gibbs updates. The values of the physical couplings are given in table~\ref{table:lambda}. To make sure there is little autocorrelation in the data, we collected a sample every $10^3$ Monte Carlo updates.
\begin{center}
\begin{table}
\begin{tabular}{ |c|c| } 
 \hline
 Coupling & Numerical Value \\
 \hline
 $\Lambda_X$ & 0.017857 \\
 \hline
 $\Lambda_T$ & 2.01273 \\ 
 \hline
\end{tabular}
\caption{\label{table:lambda}Numerical values of the physical couplings for the Ising model of Eq.~\eqref{eq:1DT}.}
\end{table}
\end{center}

\begin{figure}[ht]
 \includegraphics[scale=0.3]{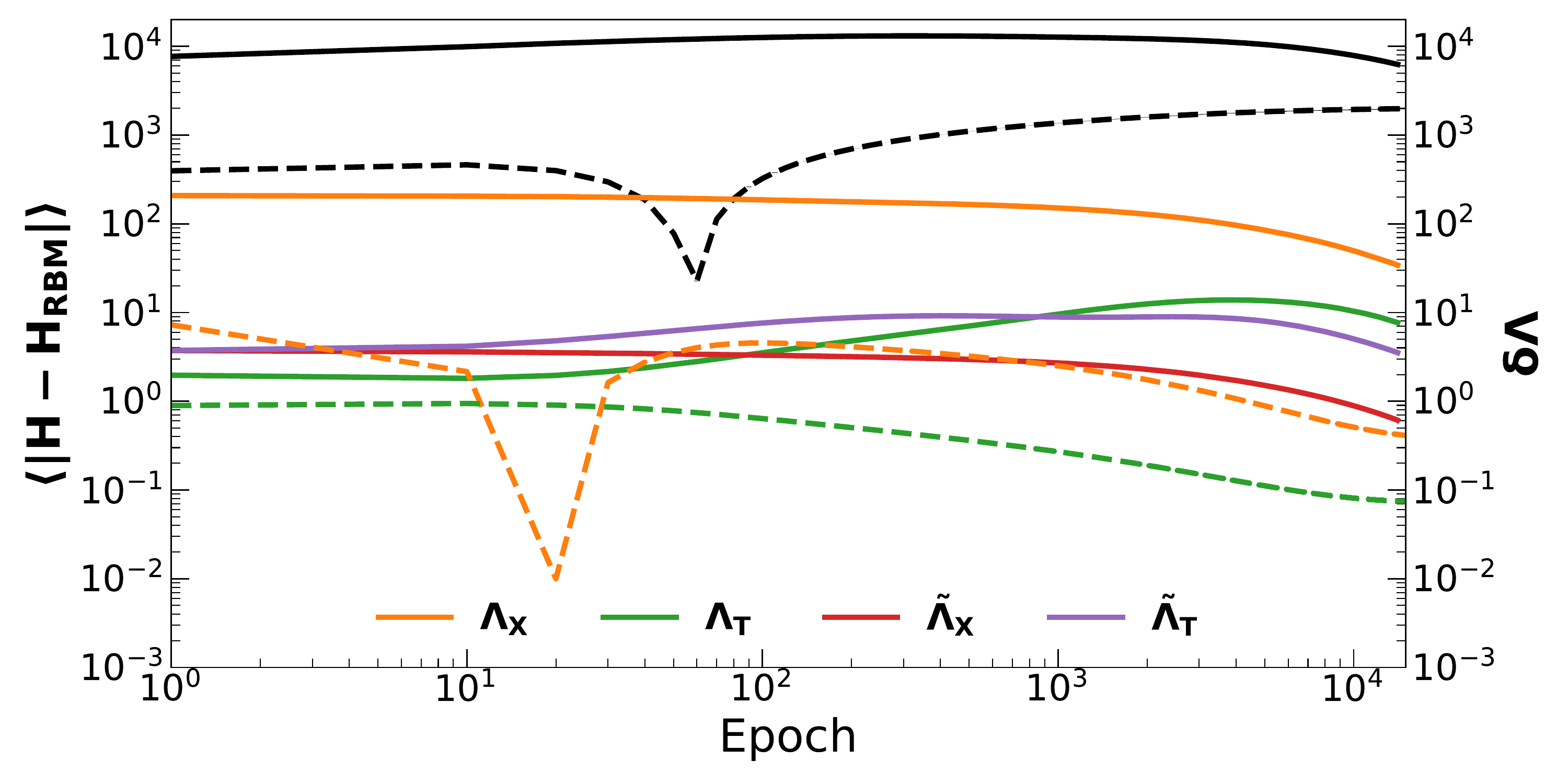}
 \caption{Training by CD of fully connected (solid lines) and sparse (dashed lines) RBM. The average energy difference in black, and relative errors in couplings $\Lambda_{X,T}$. For the fully connected RBM we also show the average values for spurious couplings denoted by $\tilde{\Lambda}$.}
\label{fig:kl_symm}
 \end{figure}
 
In Fig.~\ref{fig:kl_symm} we plot both the average error in energy per configuration and the obtained couplings as function of training epochs. One epoch consists of batch training the entire data once. 
As Fig.~\ref{fig:kl_symm} shows, while the error in the couplings decreases with training, it is about two order of magnitude lower for the sparse RBM. In addition, while in the Ising model there are only interactions among neighboring spins, the fully connected RBM includes couplings between spins that are distant and which should decrease to 0 when the training converges. These are denoted by $\tilde{\Lambda}$ in Fig.~\ref{fig:kl_symm} and are set to 0 by default for the sparse RBM which has auxiliary variables coupled only to neighboring spins. While the architecture choice greatly helped the training process, we still need to find an objective function we can easily compute and optimize for.

Having access to the probability (up to an overall norm) of a given configuration $S$ for the physical Hamiltonian, allows us to cast the optimization process as supervised learning by considering this probability as a label to be learned by the neural architecture.
In~\cite{Huang2017} the authors train the RBM by minimizing the difference in free energies, which is the difference between the Hamiltonian of the physical system and the RBM Hamiltonian for the visible layer, $\sum_{\mathcal{S}}|H(\mathcal{S})-H_{\text{RBM}}(\mathcal{S})|$.
Due to the similarity in functional form between the two, for the system studied in~\cite{Huang2017} the authors were able to engineer the one body coupling for the visible layer by directly matching to the physical system and the other parameters were optimized by minimizing the objective function. As an overall constant shift in the energy has no impact on the physics, the authors added a constant to the energies of the physical system to make them non-negative. In general, one is not able to perform a direct matching like in~\cite{Huang2017}, so we opted to implement the same objective function and training procedure, but we do not engineer any RBM parameter. We trained the RBM using various constant energy shifts and the results displayed here are the best ones obtained, for which no energy shift was used. 
\begin{figure}[ht]
 \includegraphics[scale=0.3]{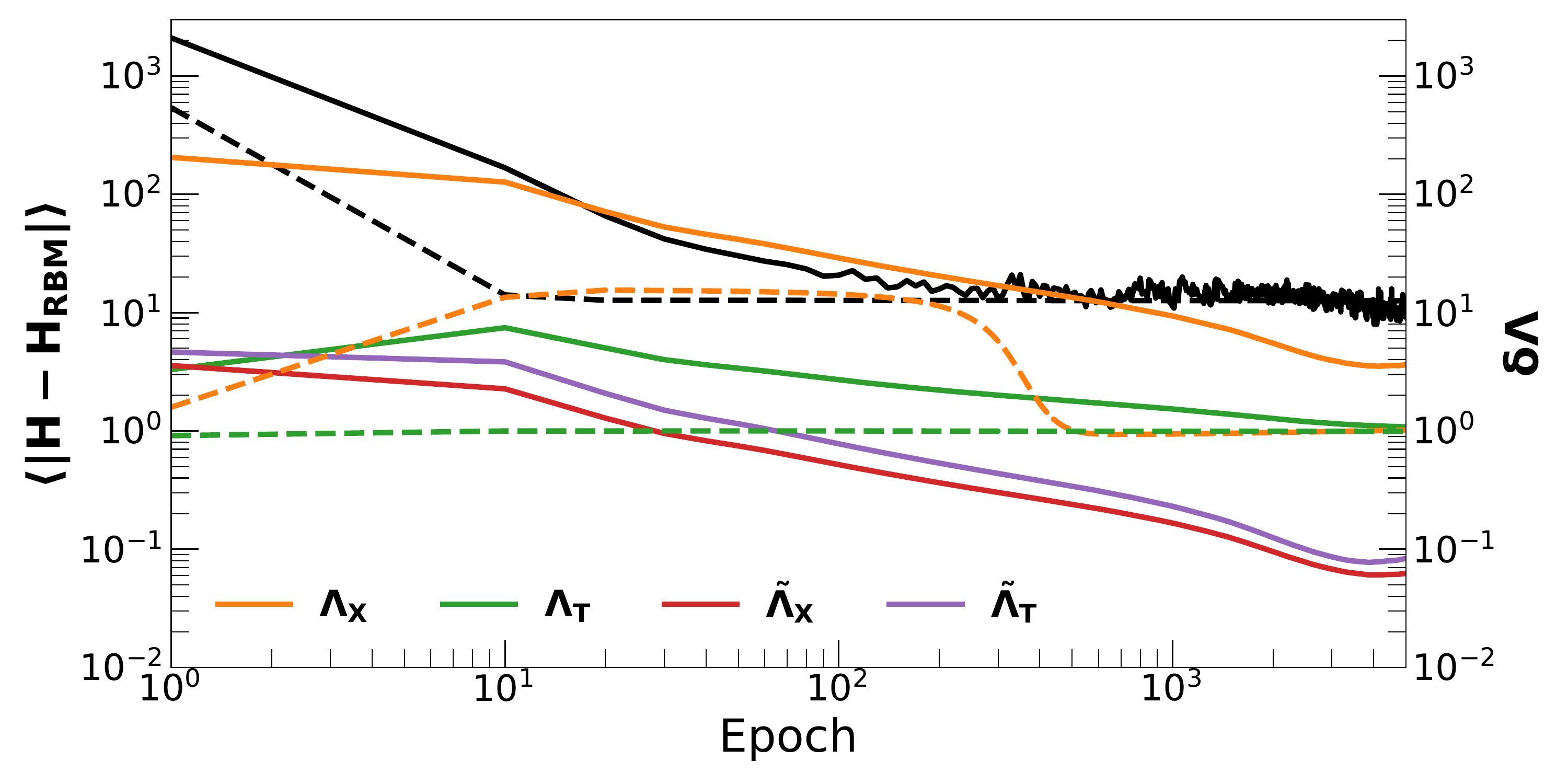}
 \caption{Training by minimizing the difference in energy of fully connected (solid lines) and sparse (dashed lines) RBM. The average energy difference in black, and relative errors in couplings $\Lambda_{X,T}$. For the fully connected RBM we also show the average values for spurious couplings denoted by $\tilde{\Lambda}$.}
\label{fig:lgpdiff_symm}
 \end{figure}
 
 The training in Fig.~\ref{fig:lgpdiff_symm} is faster than CD training in Fig.~\ref{fig:kl_symm} as the average energy difference decreases rapidly until it reaches a steady value. In addition, spurious interactions $\tilde{\Lambda}_{X,T}$ decrease rapidly as well. However, the errors in the couplings $\Lambda_{X,T}$ do not decrease to 0. To improve convergence using this objective function, one would need to also optimize for the constant energy difference, either through grid search through all possible values or by other means. We recall that the constant energy difference is related to an overall normalization factor in the RBM which can be computed exactly only by summing the probabilities of all configurations. In our experiment there are quite many of them, $2^{L_X \times L_T}\gtrsim 10^{236}$, and systems of physical interest are typically much larger.
 
To remove any dependence on the normalization factors, we introduce a new objective function which we call relative importance
\begin{equation}
\begin{split}
 \alpha&(P, P_{\text{RBM}})= \sum_{\mathcal{S}\epsilon \pi_{\mathcal{S}}} \sum_{\mathcal{S}'\epsilon \pi_{\mathcal{S}'}} \left|\ln \left(\frac{P(\mathcal{S'})P_{\text{RBM}}(\mathcal{S})}{P_{\text{RBM}}(\mathcal{S}')P(\mathcal{S})} \right) \right|\\
 =&\sum_{\mathcal{S}\epsilon \pi_{\mathcal{S}}} \sum_{\mathcal{S}'\epsilon \pi_{\mathcal{S}'}} \left| \bigtriangleup{H}(\mathcal{S},\mathcal{S}') -\bigtriangleup{H_{\text{RBM}}}(\mathcal{S},\mathcal{S}')\right|\;,
 \label{eq:ri}
 \end{split}
\end{equation}
where $\bigtriangleup{H}(\mathcal{S},\mathcal{S}')=H(\mathcal{S})-H(\mathcal{S}')$ is the difference in energy between two configurations.
We sample $\mathcal{S},\mathcal{S}'$ from two separate uniform distributions $\pi_{\mathcal{S}}\neq \pi_{\mathcal{S}'}$ defined over disjoint subsets of the training dataset $\Omega$ 
\begin{equation}
\Omega_S\cup\Omega_{S'}=\Omega\quad\Omega_S\cap\Omega_{S'}=\emptyset\;.
\end{equation}
 If the two distributions were equal, the objective function would be identically 0, so they must be chosen to be different. In our experimental setup, for simplicity, we pick a random batch in the beginning of the training to be the support of $\pi_{\mathcal{S}'}$ and the rest of the data is reserved for $\pi_{\mathcal{S}}$. The ratio inside the logarithmic function is the Metropolis-Hastings acceptance rate to apply the Monte Carlo update from $S$ to $S'$ using the RBM distribution to sample the new configuration ~\cite{Huang2017}.
\begin{figure}[ht]
 \includegraphics[scale=0.3]{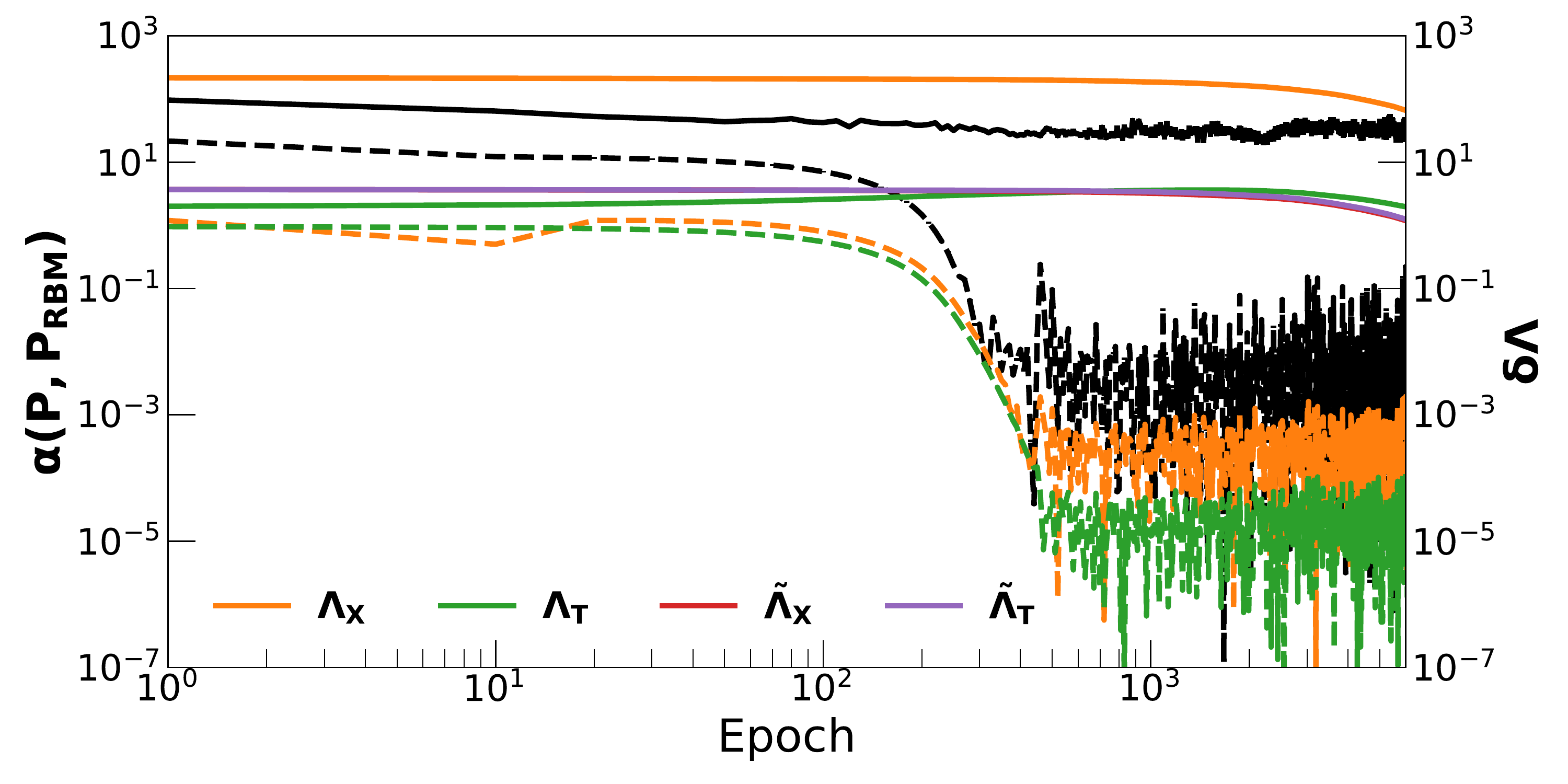}
 \caption{Training by minimizing the relative importance of fully connected (solid lines) and sparse (dashed lines) RBM. The average energy difference in black, and relative errors in couplings $\Lambda_{X,T}$. For the fully connected RBM we also show the average values for spurious couplings denoted by $\tilde{\Lambda}$.}
\label{fig:pdiff_symm}
 \end{figure}
In fig.~(\ref{fig:pdiff_symm}) we plot the error in relative importance and errors in couplings during training. Similarly to the previous training experiments, the sparse RBM produces much smaller errors than the fully connected RBM. In addition, the training for the sparse RBM is quite rapid and unlike the free energy difference, the relative importance error decreases rapidly to 0 (numerically almost 0) at the same time as the errors in the coupling decrease to 0. By combining a sparse RBM with our new objective function we have obtained complete learning and done so rather quickly in the number of training epochs.

\begin{figure}[ht]
 \includegraphics[scale=0.3]{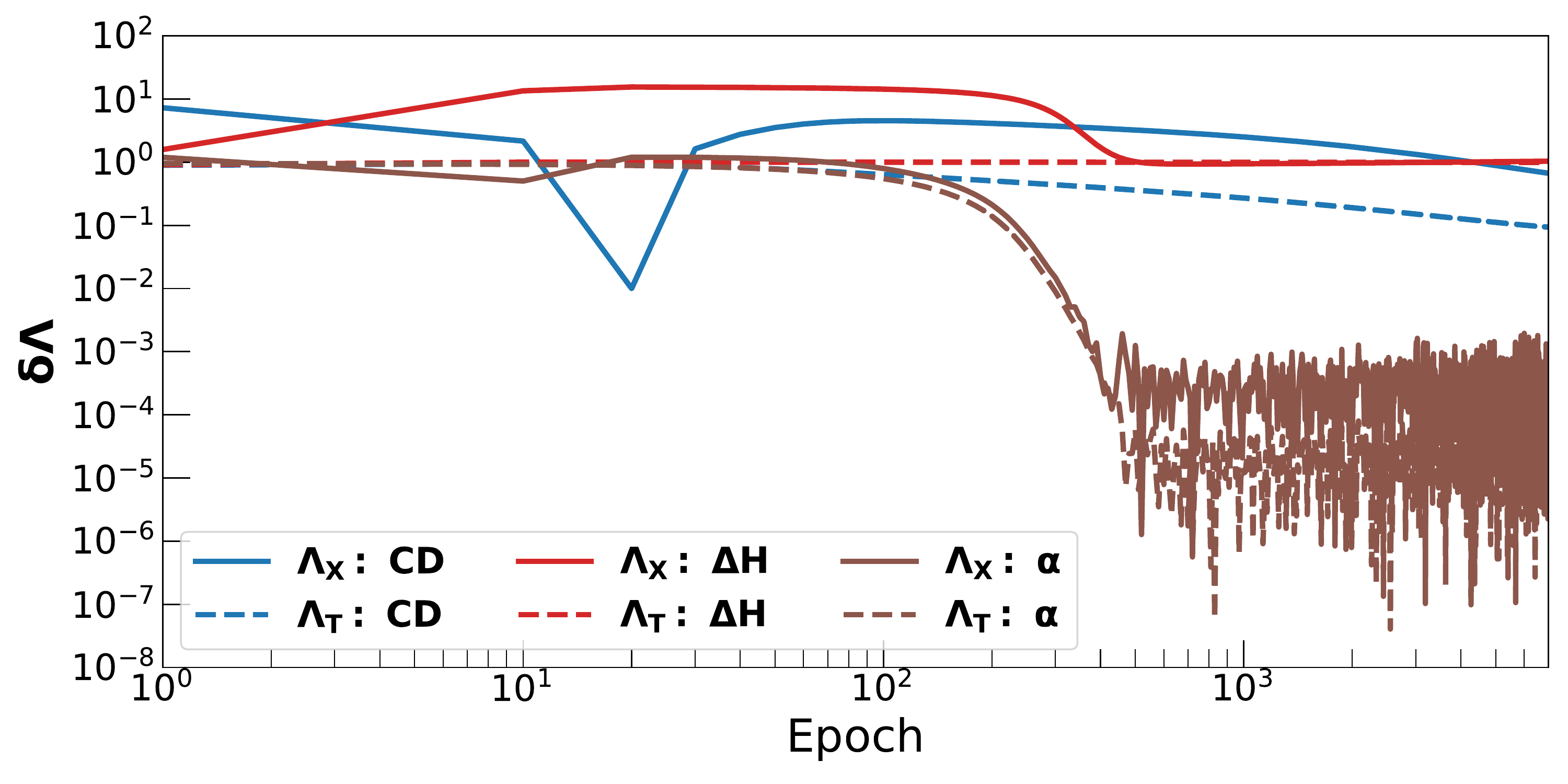}
 \caption{Relative errors in couplings $\Lambda_{X,T}$ for the sparse RBM for each of the 3 optimization methods discussed in this section.}
\label{fig:train_compare}
 \end{figure}
To compare how the 3 optimization methods perform, in fig.~\ref{fig:train_compare} we plot the relative error in the physical couplings for the sparse RBM for each of them. In all cases we have employed the same batch size and learning rate. Relative importance training converges quickly to the exact values while the two other methods either fail to converge or require training for many more epochs.

\begin{figure}[ht]
 \includegraphics[scale=0.3]{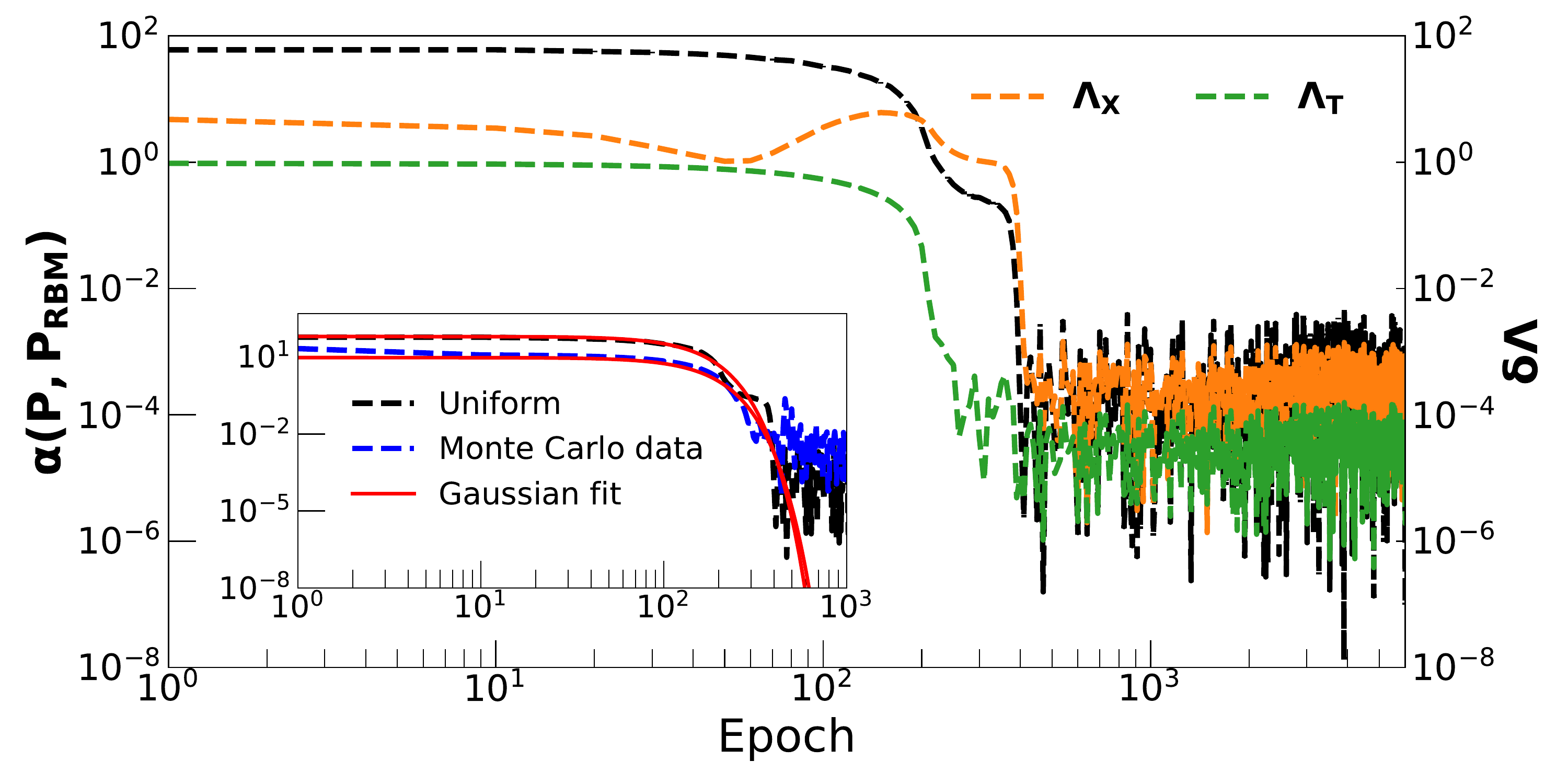}
 \caption{Training by minimizing the relative importance of the sparse RBM. The average energy difference in black, and relative errors in couplings $\Lambda_{X,T}$.}
\label{fig:pdiff_uniform}
 \end{figure}

We conclude this section with a consideration on the data used for training, which is usually assumed to represent the physical distribution one is trying to learn. However, Monte Carlo sampling used to generate the data might suffer from autocorrelation which invalidates this assumption. In~\cite{Cossu2019} this was circumvented by employing the the Swendsen-Wang cluster algorithm. In our case we had to run very long local Monte Carlo chains and discard more than $99\%$ of the samples. In cases of practical interest, one might not have any clustering algorithm available and running long Monte Carlo chains to be discarded afterwards is computationally expensive.
Since in our calculations we can evaluate the target distribution for any configuration $S$ (up to some normalization factor), the role of the training set is redundant when using the relative importance as objective function. In fact, we can even take the sampling distributions $\pi_S$ and $\pi_{S'}$ that define the relative importance to be uniform over the whole configuration space instead of just the training set $\Omega$. As a proof of concept, we perform the optimization by uniformly sampling the full configuration space with $\pi_{S'}$ having a smaller support than $\pi_{S}$. In Fig.~\ref{fig:pdiff_uniform} we plot the error in relative importance for a sparse RBM by uniform sampling the configuration space. The inset displays the objective function and Gaussian fit ($\mathcal{N}e^{-ax^2}$) for training from the data collected with local Monte Carlo updates and from uniform sampling \footnote{We found $a=6.5 \times 10^{-5}$ for uniform sampling and $a=5.3 \times 10^{-5}$ for data collected from local MC updates. These values depend on both the underlying Hamiltonian and optimization hyper parameters such as batch size and learning rate.}. As the figure shows, this procedure works quite well and the results are very similar (almost identical) to those obtained by the the sparse RBM with pre-processed data displayed in Fig.~\ref{fig:pdiff_symm}. 
 The Hamiltonian studied in this section is rather simple and mainly serves as an illustration since we can derive an exact RBM mapping of the physical couplings. In complicated settings where such mapping might not be feasible, the RBM can still be a useful approximation and the respective parameters can be obtained by the optimization procedure outlined here.

\section{Conclusion}
\label{sec:conclusion}

Computational methods based on neural networks have seen many successes in computer vision and natural language processing in recent years. In this work we used a particularly simple architecture, the Restricted Boltzmann Machine, as a tool to represent the partition function of fermionic systems with many-body interactions in terms of non-interacting fermions coupled to an external auxiliary field. This generalized Hubbard-Stratonovich transformation is obtained by introducing a hybrid quantum-classical RBM model where the visible layer is composed by quantum operators while the hidden layer represents the (classical) auxiliary fields. 

Due to their importance in Effective Field Theories in nuclear physics, we focus most of our discussion in Sec.~\ref{sec:conjecture} to contact interactions among fermions like Eq.~\eqref{eq:veft} but also discuss in detail how this can be generalized to any interaction of the form Eq.~\eqref{eq:vlocal} (see eg. the discussion at the end of Sec.~\ref{sec:2bd1}).

In order to provide support to the implementation of the RBM-based scheme we propose here, we provide details of the construction for the most common situation of two and three-body interactions in Sec.~\ref{sec:2bd1} and Sec.~\ref{sec:3bd} respectively (see also Appendix~\ref{app:General_EQ} for the general case). 

Thanks to the generality of the scheme, we expect the identities we have obtained here to prove useful in applications to Quantum Monte Carlo simulations based on auxiliary fields like Lattice-EFT and AFDMC.

An interesting application of these ideas is in finding efficient mappings of optimization problems into the native Hamiltonian of a quantum annealer as we discuss in Sec.~\ref{sec:qa}. Since the mapping is at the level of partition functions, robust techniques to extract the effective temperature will need to be used in this case.

In situations where exact representations cannot be used, we hope the optimization procedure outlined in Sec.~\ref{sec:train} could be helpful in finding useful approximations.

\begin{acknowledgments}
E.R. acknowledges the NSF N3AS Physics Frontier Cen-
ter, NSF Grant No. PHY-2020275, and the Heising-Simons
Foundation (2017-228). The work of A.R. was supported by the U.S. Department of Energy, Office of Science, Office of Advanced Scientific Computing Research (ASCR) quantum algorithm teams program, under field work proposal number ERKJ333 and by the Institute for Nuclear Theory under U.S. Department of Energy grant No. DE-FG02-00ER41132.
\end{acknowledgments}

\bibliography{references}   
% \bibliography{references2}   

\appendix
\section{Mapping between the RBM and many-body interactions}
\label{app:General_EQ}

In this appendix we provide further details on the approach we followed in producing the RBM representations for specific many-body interactions. Due to the fact that our visible layer is composed of operators, and therefore the RBM mapping is at the level of the quantum partition function, it is useful to fix a basis to work with. After the mapping is established for every element of the complete basis, we are guaranteed that it will work on any superposition state living in the same Hilbert space. A convenient basis to carry out the calculations is, not surprisingly, the eigenbasis of the operators that define the visible layer. For fermionic density operators, which are the main focus of this work, this is the Fock, or occupation number, basis for which
\begin{equation}
\hat{\rho_k}\rvert\bm{\rho}\rangle = \rho_k\rvert\bm{\rho}\rangle\quad\rho_k=0,1\;.
\end{equation}
Here the number of possible orbitals, elements in the real vector $\bm{\rho}$, is $M$ as in the main text. As we have discussed in Sec.~\ref{sec:conjecture} of the main text it is easy to see that, upon tracing out the hidden layer of the RBM, the induced Hamiltonian in the visible layer
\begin{equation}
\label{eq:rbmhama}
\begin{split}
H_{\text{rbm}}&\left( \hat{\bm{\rho}}\right)=-\ln\left(\text{Tr}_{{\bf h}}\exp\left(-F_{\text{RBM}}\left(\hat{\bm{\rho}},\bf{h}\right)\right)\right)\;,
%&=\sum_{\mu=1}^{M} B_\mu\hat{{\rho}}_\mu + \sum_{j=1}^{N_h} C_j h_j+\sum_{\mu=1}^{M}\sum_{j=1}^{N_h}W_{ij}\hat{\rho}_\mu h_j\;,
\end{split}
\end{equation}
contains all possible terms up to the maximum $M$-body interaction. This can be shown for instance using the cumulant generating function. Here we will rewrite $H_{\text{rbm}}\left( \hat{\bm{\rho}}\right)$ in a form which will simplify further manipulations. If we denote by $C(M,k)$ the set of $k$ combinations from the set of orbital indices $I=\{1,\dots,M\}$, we can express the Hamiltonian as
\begin{equation}
\label{eq:Hgeneral}
H^A_{\text{rbm}}\left( \hat{\bm{\rho}}\right) = \sum_{\mu=1}^{M} \sum_{P\in C(M,\mu)}A^{(\mu)}_P\prod_{\nu\in P}\hat{\rho}_\nu\;,
\end{equation}
which makes evident the presence of $2^M-1$ coupling terms. Note that we removed an irrelevant constant energy shift from Eq.~\eqref{eq:Hgeneral} since it only affects the overall normalization of the partition function.

We can also express the right hand side of Eq.~\eqref{eq:rbmhama} explicitly in terms of the RBM parameters, for a single hidden variable we find
\begin{equation}
H^B_{\text{rbm}}\left( \hat{\bm{\rho}}\right) = -\ln\left(\sum_{h=0}^{\mathcal{K}-1}e^{ h \left(C + \sum_{\mu=1}^M W_{\mu} \hat{\rho}_{\mu} \right) }\right)
\end{equation}
where we set the visible biases ${\bf B}$ to 0 as it is easy to reintroduce them if need be.
As discussed at the beginning of this section, we will now match these two energy functionals $H^A_{\text{rbm}}$ and $H^B_{\text{rbm}}$ on the states that span the fermionic Fock space. This allows us to replace the vector of operators $\hat{\bm{\rho}}$ with a vector of eigenvalues $\bm{\rho}$. Furthermore, if we want the RBM to represent the physical system exactly, the two energy functionals can only differ by a constant. In order to take this possible difference into account, we introduce the shift $A^{(0)}$ in Eq.~\eqref{eq:Hgeneral} and fix it's value by performing the matching in the vacuum state $\rvert0\rangle$. This immediately yields
\begin{equation}
A^{(0)}=-\ln\left(\sum_{h=0}^{\mathcal{K}-1}\exp \left( h C  \right)\right)\;.
\end{equation}
The matching condition becomes now
\begin{widetext}
\begin{equation}
 \begin{split}
\sum_{\mu=1}^{M} \sum_{P\in C(M,\mu)}A^{(\mu)}_P\prod_{\nu\in P}\rho_\nu =& -\ln \left(\frac{\sum_{h=0}^{\mathcal{K}-1}\exp \bigg( h \big(C +  \sum_{\mu=1}^M W_{\mu} \rho_{\mu} \big) \bigg)}{
  \sum_{h=0}^{\mathcal{K}}\exp \left( h C  \right)}\right)
 \end{split}
 \label{eq:HBlog} 
\end{equation}
\end{widetext}
where for simplicity we have subtracted the constant term $A^{(0)}$ on both sides. For practical purposes we can use this compact relation to construct a simple linear system for the unknown induced couplings $A^{(\mu)}_P$ using the remaining $2^M-1$ basis states (note that for the applications discussed in the main text $M$ is usually not large) and considering the right hand side as fixed for a given choice of RBM parameters. If we want a specific value for the some induced coupling we can then use the solution to this simple linear system (which is guaranteed to be non-singular) together with a root-finding algorithm to determine a possible solution for the RBM parameters that satisfy the constrain. This is the procedure we used throughout our work.

In the next subsection we show in more detail this procedure for the physically relevant cases of the two and three body forces we mention in the main text. This procedure can be repeated for higher many-body forces, and at each step, we can represent the lower order couplings in term of RBM parameters from the identities already found for them.

Similar relations can be obtained when the visible layer is composed by (for instance) spin operators by ensuring that the matching condition is enacted for every state in the corresponding eigenbasis. For Pauli matrices this will lead to the change $\rho_i\to\sigma_i=\pm1$.

\subsection{Derivation of RBM mapping for two and three body interactions}
\label{app:23bd}
Having set up the system of linear equations the allows one to solve for the many-body couplings in the previous section, we proceed to derive the identities presented in the main text for the two and three body interactions in a way that makes easy the generalization to arbitrary categorical hidden variable $h=\{0,\dots,\mathcal{K}-1\}$.

We start by denoting by $Z_2$ the general partition function with one two body interactions
\begin{equation}
\label{eq:twobpart}
 Z_2=\exp\left(-A^{(2)}{\rho}_1 {\rho}_2-A^{(1)}_{1} {\rho}_1 -A^{(1)}_{2} {\rho}_2\right)\;.
\end{equation}
We can express the equation that maps the physical system to an RBM with free energy given by Eq.~\eqref{eq:H2bd} as
\begin{equation}
 \begin{split}
  Z^{(2)}_{\text{rbm}} = \mathcal{N}_2 \sum_{h=0}^{\mathcal{K}-1} e^{-h \left(W_{1} {\rho}_1 + W_{2} {\rho}_2 + C\right)}.\\
  \end{split}
\end{equation}
where $\mathcal{N}_2$ is an overall normalization factor. As discussed above, the matching condition Eq.~\eqref{eq:HBlog} can be converted to a system of linear equations. The result is:
\begin{equation}
 \begin{split}
\begin{pmatrix}
 0 & 0 & 1 \\
 0 & 1 & 0 \\
 1 & 1 & 1 \\
\end{pmatrix}
\begin{pmatrix}
 A^{(2)} \\
 A^{(1)}_{1} \\
 A^{(1)}_{2} \\
\end{pmatrix}
 =
\begin{pmatrix}
 L^{(2)}(0,1) \\
 L^{(2)}(1,0) \\
 L^{(2)}(1,1) \\
\end{pmatrix}
 \end{split}
 \label{eq:2bd_generalv2}
\end{equation}
where the entries of the right hand side vector are
\begin{equation*}
\begin{split}
L^{(2)}( {\rho}_1, {\rho}_2)&=-\ln \left(\sum_{h=0}^{\mathcal{K}-1}\exp \left( -hC -h\sum_\mu^2 W_\mu {\rho}_\mu \right)\right)\\
&+\ln \left(\sum_{h=0}^{\mathcal{K}-1}\exp \left(- h C \right) \right)\;.
\end{split}
\end{equation*}
By solving this linear system we can find that the general expression for both the one body term
\begin{equation}
\label{eq:onebmap}
A^{(1)}_\mu = -\ln\left(\frac{\sum_{h=0}^{\mathcal{K}-1} \exp\left(-h\left(C+W_\mu\right)\right)}{\sum_{h=0}^{\mathcal{K}-1} \exp\left(-hC\right)}\right)\;,
\end{equation}
and the two body coupling is obtained as
\begin{equation}
\label{eq:twobmap}
\begin{split}
A^{(2)} = &\;\; L^{(2)}(1,1)-L^{(2)}(1,0)-L^{(2)}(0,1)\\
=&-\ln\left(\frac{\sum_{h=0}^{\mathcal{K}-1} \exp\left(-h\left(C+W_1+W_2\right)\right)}{\sum_{h=0}^{\mathcal{K}-1} \exp\left(-hC\right)}\right)\\
&-A^{(1)}_1-A^{(1)}_2 \;.
\end{split}
\end{equation}
Note that the three parameters ($W_1,W_2,C$) define a hypersurface of equivalent RBMs corresponding to the same two body coupling and different one body counter terms.

We now turn to the more involved three-body case, as we will see the structure of the mapping between physical and RBM coupling follows a similar pattern to the two body case. Let's first introduce, in analogy to Eq.~\eqref{eq:twobpart}, the three body partition function
\begin{equation}
\label{eq:3bpart}
 Z_3=e^{-A^{(3)}{\rho}_1 {\rho}_2{\rho}_3- \sum_{\mu<\nu}A^{(2)}_{\mu\nu} {\rho}_\mu{\rho}_\nu - \sum_\mu A^{(1)}_{\mu} {\rho}_\mu}\;.
\end{equation}
The relation between physical and RBM parameters for three body case is now given by
\begin{equation}
 \begin{split}
Z_3 = Z_{\text{rbm}}^{(3)}= \mathcal{N}_3 \sum_{h=0}^{\mathcal{K}-1} e^{-h \left(W_{1} \rho_1+ W_{2} \rho_2 + W_3 \rho_3 + C\right)}\;.\\
  \end{split}
\end{equation}
with $\mathcal{N}_3$ an overall normalization factor. As before we can convert this to a system of linear equations,
\begin{equation}
\begin{pmatrix}
 0 & 0 & 0 & 0 & 0 & 0 & 1 \\
 0 & 0 & 0 & 0 & 0 & 1 & 0 \\
 0 & 0 & 0 & 1 & 0 & 1 & 1 \\
 0 & 0 & 0 & 0 & 1 & 0 & 0 \\
 0 & 0 & 1 & 0 & 1 & 0 & 1 \\
 0 & 1 & 0 & 0 & 1 & 1 & 0 \\
 1 & 1 & 1 & 1 & 1 & 1 & 1 \\
\end{pmatrix}\begin{pmatrix}
 A^{(3)} \\
 A^{(2)}_{12} \\
 A^{(2)}_{13} \\
 A^{(2)}_{23} \\
 A^{(1)}_1\\
 A^{(1)}_2\\
 A^{(1)}_3\\
\end{pmatrix}=\begin{pmatrix}
L^{(3)}(0,0,1) \\
 L^{(3)}(0,1,0) \\
 L^{(3)}(0,1,1) \\
 L^{(3)}(1,0,0) \\
 L^{(3)}(1,0,1) \\
 L^{(3)}(1,1,0) \\
 L^{(3)}(1,1,1) \\
\end{pmatrix}
 \label{eq:3bd_generalv2}
 \end{equation}
where, similarly to before, we have defined
\begin{equation*}
\begin{split}
 L^{(3)}( {\rho}_1, {\rho}_2,\rho_3)&=-\ln\left(\sum_{h=0}^{\mathcal{K}-1}\exp \left(- h C -h \sum_{\mu=1}^3W_{\mu} {\rho}_{\mu}  \right) \right)\\
 &-\ln \left( \sum_{h=0}^{\mathcal{K}-1}\exp \left(- h C \right) \right)\;.
\end{split}
\end{equation*}
The solution for the one-body couplings is the same as Eq~\eqref{eq:onebmap}, while the two body term is generalized to
\begin{equation}
\label{eq:twobmap2}
\begin{split}
A^{(2)}_{\mu\nu} = &-\ln\left(\frac{\sum_{h=0}^{\mathcal{K}-1} \exp\left(-h\left(C+W_\mu+W_\nu\right)\right)}{\sum_{h=0}^{\mathcal{K}-1} \exp\left(-hC\right)}\right)\\
&-A^{(1)}_\mu-A^{(1)}_\nu \;.
\end{split}
  \end{equation}
We can now move to the solution for three body term which reads
\begin{equation}
\label{eq:threebdmap}
\begin{split}
A^{(3)} = &-\ln\left(\frac{\sum_{h=0}^{\mathcal{K}-1} \exp\left(-h\left(C+\sum_{\mu=1}^3 W_\mu\right)\right)}{\sum_{h=0}^{\mathcal{K}-1} \exp\left(-hC\right)}\right)\\
&-\sum_{\mu<\nu}A^{(2)}_{\mu\nu}-\sum_\mu A^{(1)}_\mu \;,
\end{split}
  \end{equation}
and, again, the four parameters ($W_1,W_2,W_3,C$) define a hypersurface of equivalent RBMs.
Note that the sums that appear in the definition of the physical couplings can be summed directly using standard relations for geometric progressions as we did for the two body case in Eq.~\eqref{eq:2bmap_cat_two} and Eq.~\eqref{eq:3bmap_cat} of the main text. We will make use of this in the next section.

\section{Derivation of the bounds for general categorical hidden units}
\label{app:categorical}

In this section we provide the proof for the two lower bounds Eq~\eqref{eq:2blbound} and Eq.~\eqref{eq:3blbound} presented in the main text to explain the rate of growth of the induced couplings with the maximal value $\mathcal{K}$ that defines the range of the categorical auxiliary variable $h\in\{0,\dots,\mathcal{K}-1\}$.

Given the fact that for a given target value of the physical couplings like $A_{\mathcal{K}}^{(2)}$ and $A_{\mathcal{K}}^{(3)}$ there is a continuous space of solutions for the RBM parameters, in this section we will restrict the discussion to just a one-parameter subspace for convenience. For the two-body case we choose the RBM parameters $(W_1,W_2.C)$ according to the following
\begin{itemize}
 \item for attractive interactions $A^{(2)}_\mathcal{K}<0$ we take
\begin{equation}
\label{eq:onebpar_aa}
W_1=W_2=-C=\alpha_2>0\;,
\end{equation}
 \item for repulsive interactions $A^{(2)}_\mathcal{K}>0$ we take
\begin{equation}
\label{eq:onebpar_ra}
W_1=-W_2=\alpha_2>0\quad C=0\;,
\end{equation}
\end{itemize}
while for the three-body case we take
\begin{itemize}
 \item for attractive interactions $A^{(3)}_\mathcal{K}<0$ we take
\begin{equation}
\label{eq:twobpar_aa}
W_1=W_2=W_3=\alpha_3>0\quad C=-2\alpha_3
\end{equation}
 \item for repulsive interactions $A^{(3)}_\mathcal{K}>0$ we take
\begin{equation}
\label{eq:twobpar_ra}
W_1=W_2=W_3=\alpha_3>0\quad C=-\alpha_3\;.
\end{equation}
\end{itemize}
Note that, as we mention in the main text, these parameterization choices do not have any particularly useful property other than being able to generate the desired couplings with the adjustment of a single real parameter $\alpha$. Depending on the particular application one is interested in, different choices might be closer to optimal but this will need  to be assessed on a case-by-case basis.

Using the general expressions found in the previous section, Eq.~\eqref{eq:onebmap} and Eq.~\eqref{eq:twobmap}, together with the choice of parameterization for attractive interactions Eq.~\eqref{eq:onebpar_aa}, we find that the induced one-body couplings are all equal to
\begin{equation}
\label{eq:inda1fortwb}
 \begin{split}
 A^{(1)}_{\mathcal{K}}=&-\ln(\mathcal{K})+\ln\left(\frac{1-\exp\left(\mathcal{K}\alpha_2\right)}{1-\exp\left(\alpha_2\right)}\right)\;,
 \end{split}
\end{equation}
while we find for the two-body term
\begin{equation}
\label{eq:inda2fortwb}
\begin{split}
 A^{(2)}_{\mathcal{K}} &= \left(\mathcal{K}-1\right)\alpha_2+2\ln\left(\mathcal{K}\right)-2\ln\left(\frac{1-e^{\mathcal{K}\alpha_2}}{1-e^{\alpha_2}}\right)\\
 &\lesssim -\left(\mathcal{K}-1\right)\alpha_2+2\ln\left(\mathcal{K}\right)\;,
\end{split}
\end{equation}
where we used the fact that for $\alpha_2>0$ we have
\begin{equation}
\label{eq:ineqA}
\ln\left(\frac{1-e^{\mathcal{K}\alpha_2}}{1-e^{\alpha_2}}\right)\geq\left( \mathcal{K}-1\right)\alpha_2\;.
\end{equation}

In the case of a repulsive interaction, the parameterization Eq.~\eqref{eq:onebpar_ra} leads to the following one-body terms
\begin{equation}
 \begin{split}
  A^{(1)}_1=&- \ln\left(\frac{\exp\left(-\mathcal{K}\alpha_2\right)-1}{\exp\left(-\alpha_2\right)-1}\right)+\ln(\mathcal{K})\\
  A^{(1)}_2=& -\ln\left(\frac{\exp\left(\mathcal{K}\alpha_2\right)-1}{\exp\left(\alpha_2\right)-1}\right)+\ln(\mathcal{K})\\
  %=&\ln\left(\frac{\exp\left((\mathcal{K}+1)W^{(2)}\right)-1}{\exp\left(W^{(2)}\right)-1}\right)-\mathcal{K}W^{(2)}-\ln(\mathcal{K}+1)\\
  =&\;\;A^{(1)}_1-\left(\mathcal{K}-1\right)\alpha_2\;.
 \end{split}
\end{equation}
For the induced two-body coupling, we first note that the first term in Eq.~\eqref{eq:twobmap} vanishes when taking the parameterization Eq.~\eqref{eq:onebpar_ra}. The final result reads
\begin{equation}
\begin{split}
A^{(2)}_{\mathcal{K}}&=-A^{(1)}_1-A^{(1)}_2 = -2A^{(1)}_2-\left(\mathcal{K}-1\right)\alpha_2\\
&=\left(1-\mathcal{K}\right)\alpha_2-2\ln\left(\mathcal{K}\right)+2\ln\left(\frac{1-e^{\mathcal{K}\alpha_2}}{1-e^{\alpha_2}}\right)\\
&\gtrsim \left(\mathcal{K}-1\right)\alpha_2-2\ln\left(\mathcal{K}\right)\;.
\end{split}
\end{equation}
These results can then be summarized in a single lower bound for the magnitude of the two body coupling
\begin{equation}
 |A^{(2)}_{\mathcal{K}}| \gtrsim \left(\mathcal{K}-1\right)\alpha_2-2\ln\left(\mathcal{K}\right)\;.
 \label{eq:2bdc_bound}
\end{equation}
As we mention in the main text, Eq.~\eqref{eq:2bdc_bound} shows that we have an almost linear dependence of the magnitude of the two body coupling on the maximal value of the auxiliary variable. 

Next, we turn our attention to the three body coupling. Using the parameterization for the attractive case Eq.~\eqref{eq:twobpar_aa} we find for the one body terms are all equal to
\begin{equation}
\begin{split}
A^{(1)}_{\mathcal{K}} &= -\ln\left(\frac{1-e^{\mathcal{K}\alpha_3}}{1-e^{\alpha_3}}\right)+\ln\left(\frac{1-e^{2\mathcal{K}\alpha_3}}{1-e^{2\alpha_3}}\right)\\
&=\ln\left(\frac{1+\exp\left(\mathcal{K}\alpha_3\right)}{1+\exp\left(\mathcal{K}\alpha_3\right)}\right)\;,
\end{split}
\end{equation}
and similarly for the two body terms we find
\begin{equation}
\begin{split}
A^{(2)}_{\mathcal{K}} &=-\ln(\mathcal{K})+\ln\left(\frac{1-e^{2\mathcal{K}\alpha_3}}{1-e^{2\alpha_3}}\right)-2A^{(1)}_{\mathcal{K}}\;,
\end{split}
\end{equation}
for all pair interactions. The final result for the induced three body term in the attractive case is then
\begin{equation}
 \begin{split}
A^{(3)}_{\mathcal{K}}&=3\ln\left(\mathcal{K}\right)+\left(\mathcal{K}-1\right)\alpha_3 \\
&\;\;-3\ln\left(\frac{1-e^{\mathcal{K}\alpha_3}}{1-e^{\alpha_3}}\right)+\ln\left(\frac{1+e^{\mathcal{K}\alpha_3}}{1+e^{\alpha_3}}\right)\;.
 \end{split}
\end{equation}

Moving now to the repulsive case, the parameterization Eq.~\eqref{eq:twobpar_ra} implies that the one and two body couplings are the same as what we found for the attractive two-body case in Eq.~\eqref{eq:inda1fortwb} and Eq.~\eqref{eq:inda2fortwb} but with $\alpha_2$ replaced with $\alpha_3$. The three-body interaction is instead given by
\begin{equation}
 \begin{split}
A^{(3)}_{\mathcal{K}}&=-3\ln\left(\mathcal{K}\right)-\left(\mathcal{K}-1\right)\alpha_3 \\
&\;\;+3\ln\left(\frac{1-e^{\mathcal{K}\alpha_3}}{1-e^{\alpha_3}}\right)-\ln\left(\frac{1+e^{\mathcal{K}\alpha_3}}{1+e^{\alpha_3}}\right)\;,
 \end{split}
\end{equation}
and together with the result for the attractive case we have
\begin{equation}
 \begin{split}
\left|A^{(3)}_{\mathcal{K}}\right|&=-3\ln\left(\mathcal{K}\right)-\left(\mathcal{K}-1\right)\alpha_3 \\
&\;\;+3\ln\left(\frac{1-e^{\mathcal{K}\alpha_3}}{1-e^{\alpha_3}}\right)-\ln\left(\frac{1+e^{\mathcal{K}\alpha_3}}{1+e^{\alpha_3}}\right)\\
&\gtrsim \left(\mathcal{K}-1\right)\alpha_3-3\ln\left(\mathcal{K}\right)\;.
 \end{split}
\end{equation}
To obtain the lower bound, we used the fact that 
\begin{equation}
\ln\left(\frac{1-e^{\mathcal{K}\alpha_3}}{1-e^{\alpha_3}}\right)-\ln\left(\frac{1+e^{\mathcal{K}\alpha_3}}{1+e^{\alpha_3}}\right)\geq0\;,
\end{equation}
together with the inequality Eq.~\eqref{eq:ineqA}. This concludes the proof for the bounds referenced in the main text.

\section{General quantum operators in the visible layer}
\label{sec:general_int}

The RBM network structure in our work is a hybrid quantum-classical architecture with quantum operators in the visible layer and classical 
categorical variables in the hidden layer. Up to now our main focus has been the fermionic density operator $\hat{\rho}$, and we have also provided an example for Pauli matrices in Eq.~\eqref{eq:2bdpauli}. In the first case, the RBM mapping was made possible due to the idempotency of the operator, $\hat{\rho}^2=\hat{\rho}$, and in the second case the Pauli matrix is involutory, $\sigma^2=I$. These properties are special cases of operators with generalized idempotency 
\begin{equation}
 \hat{O}^R=\hat{O},
 \label{eq:gen_idempotent}
\end{equation}
for some integer $R>1$ (for Pauli matrices $R=3$). Due to Eq.~\eqref{eq:gen_idempotent}, the highest exponent present in the Taylor series expansion of the cumulant generating function  in Eq.~\eqref{eq:binary_cgf_exp} is bounded above, $k_\mu \leq R-1$, effectively truncating the order of operator products that can be generated by the RBM.
Note that for more general operators not satisfying the idempotency condition of Eq.~\eqref{eq:gen_idempotent}, the induced operators may not have any particularly useful structure to be exploited to match to some target partition function. However, this is not a problem since, as already mentioned in the main text, any operator on a finite Hilbert space can be represented by direct products of Pauli matrices. The general idempotency described here, while not necessary, could be useful if present in a given Hamiltonian.
\end{document}